\documentclass[twocolumn,prb,aps,longbibliography,showpacs,
superscriptaddress,10pt]{revtex4-1}

\usepackage{amsfonts,amssymb}
\usepackage[sumlimits,intlimits]{amsmath}
\usepackage{graphics}
\usepackage{graphicx}
\usepackage{color}
\usepackage{mathrsfs}
\usepackage{textcomp}
\usepackage{verbatim}
\usepackage{hyperref}    
\usepackage{bm}          

\newcommand{\be}{\begin{equation}}
\newcommand{\ee}{\end{equation}}
\newcommand{\e}{\varepsilon}
\newcommand{\kf}{k_\text{F}}
\newcommand{\Ni}{n_\text{imp}}

\newcommand{\ltf}{\lambda_\text{TF}}

\newcommand{\ko}{K_\circ}
\newcommand{\kp}{K_\parallel}
\newcommand{\lD}{l_\text{dim}}

\def\cH{\hat{\cal H}}

\def\bB{{\bf B}}

\def\bk{{\bf k}}

\def\bn{{\bf n}}
\def\bq{{\bf q}}
\def\bp{{\bf p}}
\def\br{{\bf r}}

\def\hn{\hat n}

\def\hsigma{\hat\sigma}

\begin{document}

\title{Electron transport in nodal-line semimetals}

\author{S.V.~Syzranov}
\affiliation{Joint Quantum Institute, NIST/University of Maryland, College Park, MD 20742, USA}
\affiliation{School of Physics and Astronomy, Monash University, Victoria 3800, Australia}

\author{B. Skinner}
\affiliation{Massachusetts Institute of Technology, Cambridge, MA 02139 USA}


\date{\today}

\begin{abstract}

We study the electrical conductivity in a nodal-line semimetal with charged impurities.  The screening of the Coulomb potential in this system is qualitatively different from what is found in conventional metals or semiconductors, with the screened potential $\phi$ decaying as $\phi \propto 1/r^2$ over a wide interval of distances $r$.  This unusual screening gives rise to a rich variety of conduction regimes as a function of temperature, doping level and impurity concentration. In particular, nodal-line semimetals exhibit a diverging mobility $\propto 1/|\mu|$ in the limit of vanishing chemical potential $\mu$, a linearly-increasing dependence of the conductivity on temperature, $\sigma \propto T$, and a large weak-localization correction with a strongly anisotropic dependence on magnetic field.

\end{abstract}


\maketitle

A topological semimetal is a semimetal in which the conduction and valence bands touch at a point
or a line (the nodal line) in momentum space. The most well-known examples are Weyl\cite{Wan:WeylProp, ZHasan:TaAs, ZHasan:TaAs2, ZHasan:TaP, ZHasan:NbAs, Weng:PhotCrystWSM},
nodal-line\cite{ZrSiS:first,PbTaSe2:first}
and parabolic\cite{Shin:quadratic} semimetals. Such materials have been in the focus of researchers' attention
since their discovery several years ago, owing to the abundance of novel fundamental
phenomena predicted in them, including the chiral anomaly\cite{Parameswaran:Anomaly},
topologically protected Fermi arcs\cite{Wan:WeylProp,ZHasan:TaAs2,Weng:PhotCrystWSM},
and unconventional disorder-driven transitions\cite{Fradkin2}\footnote{See Ref.~\onlinecite{Syzranov:review} for a review}.

In the presence of disorder, topological semimetals may display a rich variety of electrical conduction regimes, depending on
the nature of quenched disorder, temperature, and doping level.  For example, these systems may
exhibit metallic or insulating transport properties or unconventional dependencies 
of the conductivity on temperature. In general, transport in topological semimetals is determined by the behaviour
of several potentially competing factors at energies near the band touching, including 
the vanishing density of states (in the disorder-free limit),
divergence of the screening length and the behaviour of the elastic scattering rate.

Although the conductivity of Weyl semimetals has received considerable attention in the literature, 
other types of semimetals have largely evaded researchers' attention. In this paper we study
the conductivity of a three-dimensional (3D) nodal-line semimetal (NLS),
where the conduction and valence bands touch along a line in momentum space,
as in the recently discovered $\text{ZrSiS}$\cite{ZrSiS:first}, HfSiS\cite{Ando:HfSiS} and $\text{PbTaSe}_2$\cite{PbTaSe2:first}
(a number of other materials have also been predicted to be NLSs\cite{BurkovHookBalents, KimKane:prediction, Xie:Ca3P2prediction, FangFu:prediction, Gan:XB6, Yu:Cu3PdN, Mullen:hyperhoneycomb, ZHasan:TlTaSe2}).
We calculate the NLS conductivity microscopically and study its dependence on temperature, doping level
and impurity concentration for realistically achievable regimes.
We demonstrate that the Coulomb interaction in a low-doped NLS 
is only partially screened and has the distance dependence $\propto1/r^2$ across a broad interval of $r$, which leads to 
qualitatively new features in conduction and 
a richer variety of transport regimes as comopared to other semimetals, semiconductors
and metals. In particular, NLSs display divergent quasiparticle mobility in
the limit of vanishing doping, linear dependence of conductivity on temperature and large weak-localisation
corrections to the conductivity with a strongly anisotropic dependence on magnetic field.

\begin{figure}[b]
	\centering
	\includegraphics[width=0.4\textwidth]{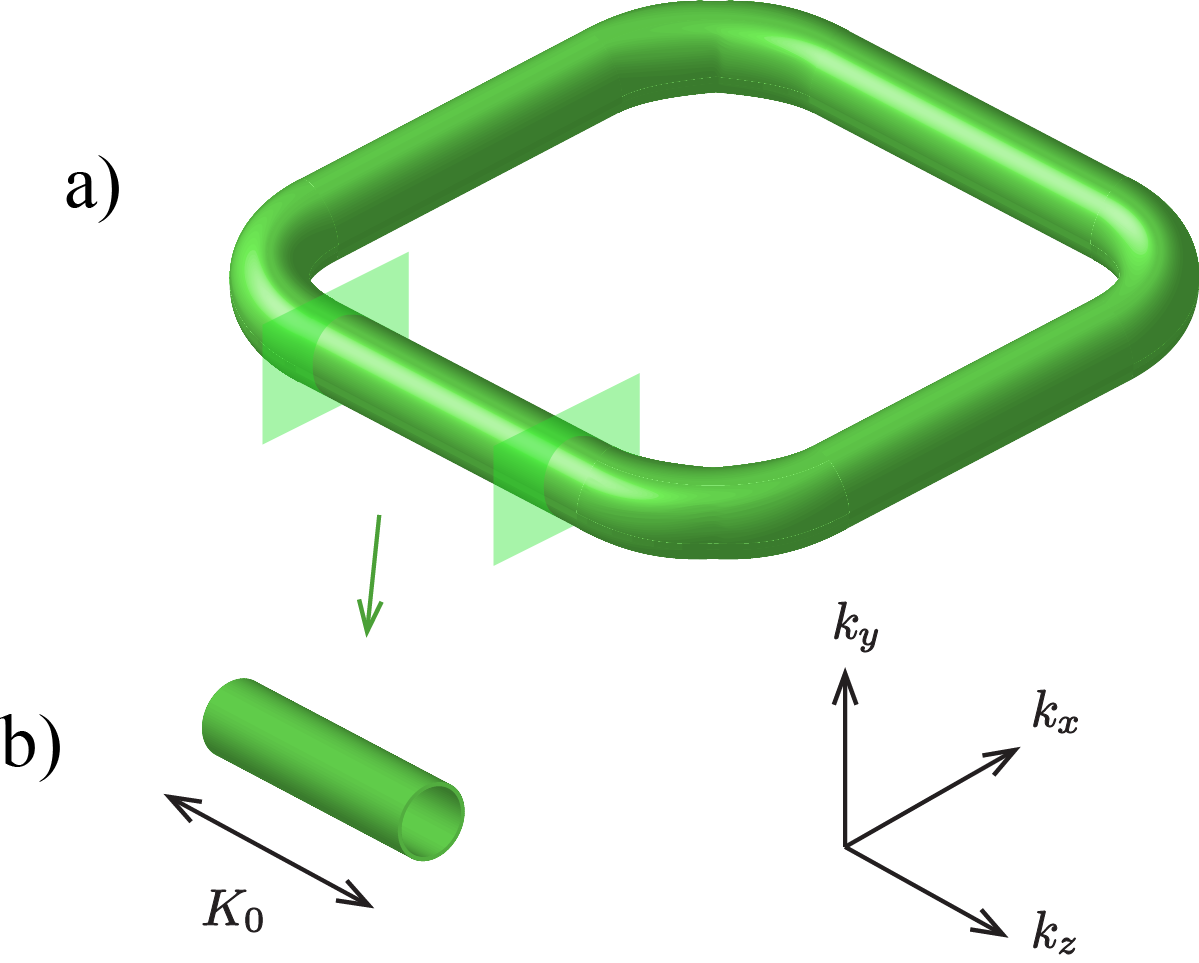}
	\caption{
	(Colour online)
	Fermi surface (FS) in a doped NLS. a) FS for a closed nodal line.
	b) A straight segment of the FS.
	For a long-range-correlated disorder potential, characteristic of Coulomb impurities,
	the momentum scattering along the nodal line is small, and the line may be approximated
	by a chain of straight segments. The total conductivity is given by the sum
	of the segment contributions.}
	\label{Fig:Tubes}
\end{figure}

{\it Model.} 
It may be assumed normally that in an undoped NLS
all points of the nodal line lie at the Fermi energy. 
In a doped material, where a finite carrier concentration is provided either by donor or acceptor impurities
or by additional pockets of states,
the Fermi surface has the shape of a tube surrounding the nodal line, as depicted in Fig.~\ref{Fig:Tubes}a.
For realistic doping levels,
the radius of the tube (the Fermi momentum) is significantly smaller than the characteristic size of the nodal
line, which is typically of order of the inverse lattice spacing.

We also assume that the dominant source of momentum scattering for quasiparticles
is provided by Coulomb impurities, as is typical for semiconductors and semimetals.  
Due to the long-range nature of the potential of these impurities, they provide only small-momentum scattering relative to the diameter of the nodal line,
so that the curvature of the nodal line may be neglected in each scattering event.
Scattering processes between opposite sides of the nodal line can also be neglected.

The Hamiltonian for a quasiparticle near a short straight segment of the nodal line is given by
\begin{align}
	\cH_0 = v(\hsigma_x k_x + \hsigma_y k_y)+\xi(k_z) +e\phi(\br),
	\label{H0}
\end{align}
where $k_x$ and $k_y$ are the transverse momentum components (hereinafter $\hbar=1$);
$\hsigma_x$ and $\hsigma_y$ are Pauli matrices in the space of a spin-$1/2$ degree of freedom
(pseudospin space);
$v$ is the velocity of transverse motion, which for simplicity is assumed to be the same in the $x$ and $y$ directions
throughout this paper;
$\xi(k_z)$ is the contribution of the longitudinal motion to the kinetic energy; and
$\phi(\br)$ is the electric potential created by charged impurities screened by electrons.

For the small quasiparticle energies under consideration the longitudinal quasiparticle velocity
may be estimated as $v_z\simeq vk_z/p_0$, and it is strongly suppressed compared to the transverse velocity $v$,
where $p_0$ is the local radius of curvature of the nodal line (in momentum space). As a result, the quasiparticle dynamics on
sufficiently short length scales is effectively 2D and is confined to the plane perpendicular to the segment of the nodal line under consideration.  

The strength of the Coulomb interaction in an NLS may be characterised by the effective fine structure constant
$
	\alpha=\frac{e^2}{\varkappa v}
$
(in Gaussian units), with $\varkappa$ being the dielectric constant. Usually $\alpha\lesssim1$ in semimetals
(see, e.g., Refs.~\onlinecite{Skinner:WeylImp} and \onlinecite{Rodionov:donorsacceptors} for estimates).
In this paper we assume for simplicity that $\alpha\ll1$, which allows one to use the linear Poisson
equation to describe the 
electrostatic potential $\phi(\br)$ created by screened charged impurities:
\begin{align}
	\varkappa\nabla^2\phi(\br)+4\pi e^2\int \Pi(\br,\br^\prime)\phi(\br^\prime)\:d\br^\prime
	=
	-4\pi e\sum_j Z_j\delta(\br-\br_j).
	\label{GenEquationPhiLin}
\end{align}
Here, $\br_j$ and $Z_je$ are the location and the charge of the $j$th impurity (for donors and acceptors
$Z_j=\pm1$, respectively) and
$\Pi(\br,\br^\prime)=-i\int_{-\infty}^{0}\langle[\hn(0,\br),\hn(t^\prime,\br^\prime)]\rangle dt^\prime$ is the
zero-frequency polarisation operator, which describes the linear response
of the local density of electrons $\hn(\br)$ to
the electrostatic potential $\phi(\br^\prime)$.

{\it Polarisation operator.} 
Due to the effectively two-dimensional short-distance dynamics of the quasiparticles,
the screening properties
of electrons near a short segment of the nodal line are related to those in
graphene\cite{Ando:polarisation,HwangSarma:GraphenePolarisation},
with the contribution to the polarisation operator given by that of 2D Dirac electrons multiplied by $g K_\parallel/(2\pi)$, where $g$ indicates the spin and valley degeneracy and $K_\parallel$ is the length of this segment in momentum space.  The full polarisation operator is given by a sum of all such straight-line-segment
contributions, since the entire nodal line may be approximated as a chain of straight-line segments.

At low temperature and chemical potential the polarisation operator $\Pi(\bq)$ is linear 
in the momentum $|\bq|$ for any direction of $\bq$.
While the constant of proportionality between $\Pi(\bq)$ and $|\bq|$ for a given direction of $|\bq|$ depends in general on the shape of the nodal 
line\footnote{These constants of proportionality between $\Pi(\bq)$ and $|\bq|$ are calculated for the case of a circular nodal line in Ref.~\onlinecite{Huh:NLS}.}, below we assume for simplicity that the full polarisation operator
is isotropic in $\bq$. For sufficiently high chemical potentials $\mu$ or temperatures $T$, 
$\Pi(\bq)$ is momentum-independent and is determined by the density of states (DoS) at energies $\sim \max (|\mu|, T)$.
Thus, the behaviour of the polarisation operator in an NLS
in the limits of high and low temperatures may be summarised as
\begin{align}
	\Pi(\bq)=
	\left\{
	\begin{array}{cc}
		-g C K_\circ |\bq|/v, 				& v|\bq|\gg T, |\mu|,\\
		-g K_\circ|\mu|/(4\pi^2 v^2), & |\mu|\gg T, v|\bq|, \\
		-g K_\circ T\ln2/(2\pi^2 v^2), & T\gg |\mu|, v|\bq|.
	\end{array}
	\right.
	\label{PolarisationOperator}
\end{align}
Here, $K_\circ$ is the length of the nodal line and $C$ is a constant of order unity, which accounts for the details of the geometry
of the nodal line.

{\it Screened impurity potential.}
At low temperature and chemical potential, the distance dependence of the screened Coulomb potential in an NLS 
is qualitatively different from that in conventional metals, dielectrics, or other semimetals. 
The Fourier transform of the 
the screened interaction is given by
$
	\phi(\bq)=\phi_0(\bq)
	\left[
	1-\Pi(\bq)\phi_0(\bq)
	\right]^{-1},
$
where $\phi_0(\bq)=4\pi e^2/(\varkappa q^2)$ describes the unscreened Coulomb interaction
and the polarisation operator $\Pi(\bq)$ is given by Eq.~(\ref{PolarisationOperator}).
\begin{figure}[htbp]
	\centering
	\includegraphics[width=0.35\textwidth]{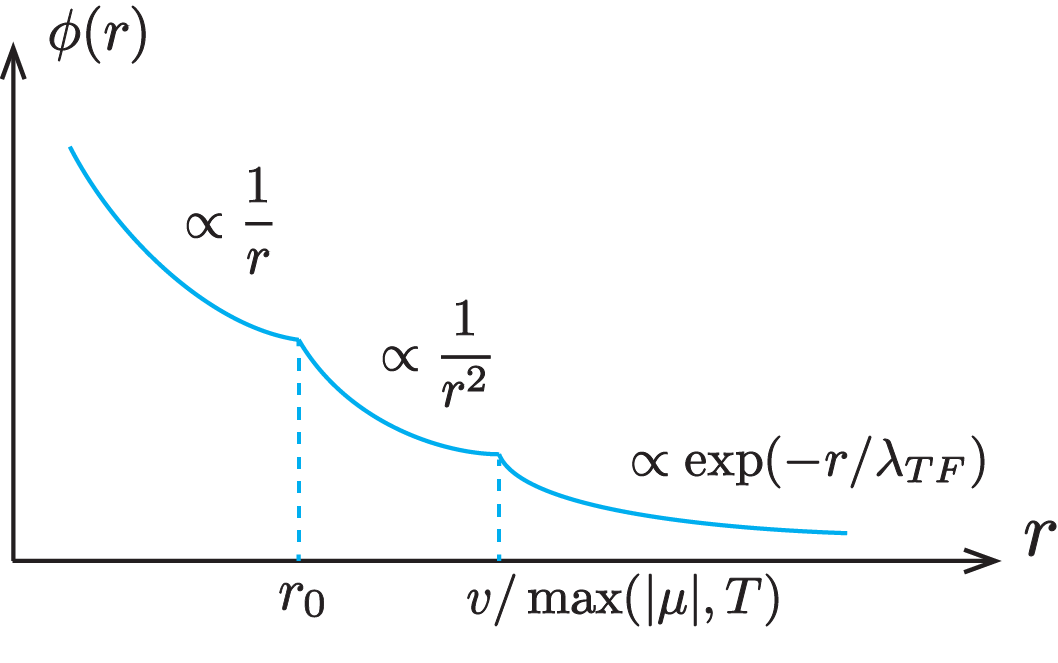}
	\caption{(Colour online) Screened Coulomb interaction as a function of distance at low
	temperatures and doping levels.}
	\label{Fig:ScreeningPlot}
\end{figure}
At short distances the interaction is unscreened, $\phi(r)\simeq e^2/(\varkappa r)$.
At distances of order of
\begin{align}
	r_0=(\alpha g K_\circ)^{-1}
	\label{r0}
\end{align}
the interaction potential
crosses over to the unconventional form
\begin{align}
	\phi(r) = \frac{e^2 r_0}{2\pi^2\varkappa C}\frac{1}{r^2}.
	\label{PhiInterm}
\end{align}
Finally, at very large distances, exceeding the characteristic wavelength $\max(|\mu|,T)/v$
of the quasiparticles in the conduction (valence) band, the polarisation operator is effectively local,
$\Pi(\br,\br^\prime)\approx -(4\pi e^2\ltf^2/\varkappa)^{-1}\:\delta(\br-\br^\prime)$, resulting in the exponentially
suppressed interaction $\phi(r)\propto \exp(-r/\ltf)$.
Here we have introduced the Thomas-Fermi (TF) screening length, given by
\begin{align}
	\ltf^{-2}=
	\left\{
	\begin{array}{cc}
		(g\alpha/\pi) \cdot K_\circ|\mu|/v, \quad & |\mu|\gg T,
		\\
		(2g\alpha/\pi)\ln2\cdot K_\circ T/v, \quad & T\gg|\mu|.
	\end{array}
	\right.
	\label{eq:lTF}
\end{align}
The dependence of the screened interaction on distance is summarised
in Fig.~\ref{Fig:ScreeningPlot}, assuming low temperature and chemical potential
($|\mu|,T\ll \alpha gv\ko$). 

For high temperatures or chemical potentials, $\max(|\mu|,T)\gg \alpha gv\ko$, the characteristic
quasiparticle wavelength
becomes shorter than the distance $r_0$, and the intermediate regime
with $\phi(r)\propto1/r^2$ (see Fig.~\ref{Fig:ScreeningPlot}) vanishes. In this case the screened
electrostatic potential is given by
\begin{align}
\phi(r) = \frac{e^2}{\kappa r}\exp(-r/\ltf)
\label{Yukawa}
\end{align}
across all distances, as in a conventional metal\cite{Abrikosov:metals}.

{\it Quasiparticle scattering.}
As mentioned in the introduction, it is possible to approximate the nodal line by a chain of straight-line segments and to consider
separately the quasiparticle transport near each segment. Due to the suppressed quasiparticle motion along
the nodal line, the transverse conductivity $\sigma_x$ of a given segment with momentum length $\kp$ significantly exceeds its
longitudinal conductivity $\sigma_z$. The sum of the conductivity from all segments, with their various orientations, is then of order of $\sigma_x \ko/K_\parallel$.

The relaxation of the momentum of quasiparticles with energy $\varepsilon$
(transverse momentum $\varepsilon/v$) near a straight segment of the nodal line
due to elastic scattering off impurities is characterised by the
transport scattering time $\tau_{\text{tr}}$. In the Born approximation,
\begin{align}
	\frac{1}{\tau_{\text{tr}}(\varepsilon)}=
	2\pi n_{\text{imp}}\int \frac{d\bp}{(2\pi)^3}|\phi(\bp-\bk)\langle\sigma_\bp|\sigma_\bk\rangle|^2
	\nonumber\\
	(1-\cos\theta_{\bk,\bp})\,\delta(kv-pv),
	\label{TranspRateGen}
\end{align}
where $n_{\text{imp}}$ is the total concentration of (donor and acceptor) impurities, 
$\bk$ is a momentum with the transverse component $\varepsilon/v$,
$\phi(\bq)$
is the Fourier-transform
of the impurity potential,
and $|\sigma_\bp\rangle$ is the pseudospin state of a quasiparticle
with momentum $\bp$ in a given (conduction or valence) band;
$|\langle\sigma_\bp|\sigma_\bk\rangle|^2=(1+\cos\theta_{\bk,\bp})/2$, with $\theta_{\bk,\bp}$ being the angle between
the transverse components of $\bk$ and $\bp$.

{\it Low doping levels.} 
At low doping, $|\mu| \ll \alpha g \ko$, and zero temperature,
the Fermi wavelength $\sim \mu/v$ significantly exceeds 
the length scale $r_0$ [Eq.~(\ref{r0})].
Thus, the scattering of quasiparticles at the Fermi surface
is determined by the $\propto 1/r^2$ tail of the impurity potential
rather than the $\propto 1/r$ core (see Fig.~\ref{Fig:ScreeningPlot}).  

Using Eqs.~(\ref{PhiInterm}) and (\ref{TranspRateGen}), we find the transport scattering time in this regime to
be (see Supplemental Material for details)
\begin{align}
	\tau_{\text{tr}}=\gamma_{\text{tr}}g^2\ko^2/(n_{\text{imp}} v),
	\label{TransportTimeLow}
\end{align}
where $\gamma_{tr}$ is a non-universal constant of order unity, which depends on the details
of the geometry of the nodal line. We note that typical scattering events have large scattering
angles $\theta_{\bp,\bk}\sim 1$, and therefore the transport scattering time (\ref{TransportTimeLow})
is of the same order as the typical time $\tau_0$
between collisions (the elastic scattering time).

For weak disorder, such that 
 $\tau_0 \mu\gg 1$,
the DoS of quasiparticles
at the Fermi energy is weakly affected by impurities, and the conductivity is dominated by the 
Drude contribution\cite{Efetov:book,AGD,Abrikosov:metals},
which takes into account quasiparticle scattering processes not involving quasiparticle interference.
The transverse Drude conductivity of the straight nodal-line segment
is given by  
\be 
\sigma_x(\mu)=\frac{1}{2}\frac{K_\parallel}{2\pi}ge^2v^2\nu_{2D}(\mu) \tau_{\text{tr}}
= \frac{\gamma_{\text{tr}}}{8\pi^2} \frac{g^3 e^2 \ko^2 \kp |\mu|}{n_{\text{imp}}v},
\label{eq:sigmaklow}
\ee
where $\nu_{2D}=\frac{|\mu|}{2\pi v^2}$ is the quasiparticle DoS (per spin per valley)
in the transverse 2D plane. As discussed above, the conductivity of the entire NLS is of order (\ref{eq:sigmaklow})
with the replacement $\kp\rightarrow \ko$.

Since the concentration of charge carriers depends quadratically on the chemical potential, $n(\mu)\propto\mu^2$,
Eq.~(\ref{eq:sigmaklow}) implies that the quasiparticle mobility $\frac{\sigma_x(\mu)}{e n(\mu)}
\propto\frac{1}{|\mu|}$
diverges as the chemical potential approaches the nodal line, even for a fixed impurity concentration. 

{\it Temperature dependence.} \;
In the regime of low doping under consideration ($|\mu| \ll \alpha g v \ko$), the conductivity strongly depends on
temperature when $T\gg|\mu|$ (here we neglect electron-phonon scattering and the interaction
between quasiparticles).  In the limit of weak disorder, 
we find for the
conductivity of a straight segment of the nodal line
\begin{align}
	\sigma_x(T\gg|\mu|)=-\int n_\text{F}^{\prime}(\varepsilon) \sigma_x(\varepsilon) d\varepsilon
	=\frac{\gamma_{\text{tr}}\ln2}{4\pi^2} \frac{g^2 e^2 \ko^2 \kp}{n_{\text{imp}}v} T,
	\label{SigmaHighT}
\end{align}
where $n_\text{F}(\varepsilon) = 1/[\exp(\varepsilon/T) + 1]$ is the Fermi distribution function
and $\sigma_x(\varepsilon)$ is given by Eq.~(\ref{eq:sigmaklow}).
The conductivity of the NLS is described approximately by Eq.~(\ref{SigmaHighT}) with the replacement $K_\parallel\rightarrow\ko$. 
The linear-in-$T$ dependency of the
conductivity
comes from thermally activated charge carriers with a linear
density of states $\nu(\varepsilon)\propto\varepsilon$ and a constant scattered rate (\ref{TransportTimeLow}). 



{\it High-doping regime.}
When the NLS is so heavily doped that $|\mu| \gg \alpha g v \ko$,
the TF screening radius becomes shorter than $r_0$ and the screened potential
is equivalent to that in a conventional metal 
[Eq.~(\ref{Yukawa})].  In this regime
the transport properties of the NLS are also similar to those of a conventional metal and, in particular, weakly dependent on the
temperature. Using Eqs.~(\ref{Yukawa}) and (\ref{TranspRateGen}) we find the
transport scattering rate in this regime\footnote{This transport scattering is equivalent to the scattering of electrons
on the surface of a 3D topological insulator with bulk impurities\cite{Skinner:TIsurface}.}
\be
	\frac{1}{\tau_{\text{tr}}(\varepsilon)} = 
	\frac{2\pi\alpha^2 v^3 n_{\text{imp}}}{\varepsilon^2}\ln\frac{|\varepsilon| \ltf}{v}.
	\label{TransportRateHigh}
\ee
At such high doping the impurities typically scatter quasiparticle momenta
by small angles $\theta\ll1$, which leads to a significantly shorter elastic scattering time
$\tau_0=(4\pi\alpha^2 \Ni v \ltf^2)^{-1}$ (see Supplemental Material for details).
The Drude conductivity of a highly-doped NLS is then given by the segment contribution
\be 
\sigma_x(\mu) = \frac{g \kp e^2 |\mu|^3}{8 \pi^3 \alpha^2 v^3 n_{\text{imp}} \ln 
\left[|\mu|/(\alpha g \ko v) \right] },
\label{eq:sigmakhigh}
\ee 
with $\kp$ replaced by a quantity of order $\ko$. 

{\it Weak localisation.}
For all doping levels, the quasiparticle dynamics on sufficiently short length scales 
are effectively 2D and are confined to the plane perpendicular to the nodal line.  One can therefore expect
significant quantum interference effects that are typical
of 2D systems\cite{Efetov:book,Gantmakher:book}.
In order to estimate the characteristic length scale $\lD$ below which these interference effects are strong,
we assume that the quasiparticles have a small longitudinal dispersion $\xi(k_z) = vk_z^2/2p_0$, \textit{e.g.}, due to
the curvature of the nodal line (in which case $p_0$ is the local radius of curvature of the line).

To estimate $\lD$, let us consider a quasiparticle whose momentum $\bk$ at time $t = 0$ lies in the $x$-$y$ plane.
Due to the small dispersion in the $z$ direction and collisions with impurities, such a quasiparticle
slowly drifts away from the $x$-$y$ plane.
During each collision with an impurity, the longitudinal velocity of the quasiparticle changes
by an amount $\delta v_z \sim \pm v k/\ko$, where we have assumed that the local radius $p_0$ of the nodal
line is of order $\ko$.
After a long time $t \gg \tau_0$, the total change in $z$ velocity is of order $\Delta v_z \sim (v k/\ko) \sqrt{t/\tau_0}$,
and the quasiparticle travels a distance $\Delta z \sim t \Delta v_z \sim (v k/\ko) \sqrt{t^3/\tau_0}$
in the $z$ direction.  When this drift length exceeds the electron wavelength $\sim k^{-1} = v/|\mu|$, the quasiparticle
can be considered to have ``escaped" its initial 2D plane of motion and no longer participates in the 2D interference
effects in this plane. The characteristic time of 2D interference can thus be estimated as
$t_{\text{dim}}\sim (v^2\ko^2\tau_0/\mu^4)^\frac{1}{3}$, which corresponds to the length
\begin{align}
	\lD = (v^4 \ko/\mu^2)^\frac{1}{3}\tau_{\text{tr}}^\frac{1}{2}\tau_0^\frac{1}{6}
	\label{l2D3D}
\end{align}
of diffusion in the initial 2D plane. 

Thus, quantum interference effects in an NLS on short distances $L< \lD$ are equivalent to those
of 2D Dirac fermions, such as electrons in graphene in a single valley, which exhibit 
 weak-antilocalisation (WAL) corrections to conductivity\cite{McCann:WL,AleinerEfetov}.
On larger length scales $L> \lD$, the classical trajectories of the quasiparticles are essentially 3D
and the interference between them is suppressed. Thus, the
conductivity of the NLS receives a WAL correction
\begin{align}
	\delta\sigma_{WL}=\frac{ge^2 \ko}{(2\pi)^3}\times
	\left\{
	\begin{array}{cc}
		\ln\left[{\min(l_\phi,l_{\text{dim}})}/(v\tau_{\text{tr}})\right], &
		B\ll B_0,
		\\
		\tilde\gamma_B(\bn_\bB)\cdot\ln[l_B/(v\tau_{\text{tr}})], & B\gg B_0,
	\end{array}
	\right.
	\label{WAL}
\end{align}
where $B_0=\frac{c}{ev\tau_{\text{tr}}\min(l_\phi,l_{\text{dim}})}$ is the characteristic value of magnetic field
at which the WAL is affected by the field, $\tilde\gamma_B(\bn_\bB)<1$ is a non-universal coefficient depending on the geometry of the nodal line and the direction $\bn_\bB=\bB/B$ of the magnetic field,
$l_\phi$ is the dephasing length,
and $l_B=\frac{c}{ev\tau_{\text{tr}} B}$.

The first line in Eq.~(\ref{WAL}) describes
the conventional WAL corrections
due to the interference of 2D Dirac fermions\cite{McCann:WL,AleinerEfetov};
the second line describes the partial suppression of WAL by magnetic field.
Because the correction is dominated by effectively 2D interference effects,
the magnetic field $\bB$ does not suppress the interference near parts of the nodal line perpendicular to $\bB$
(here we neglect the much weaker effect of magnetic field on the motion along the nodal line), and
therefore the suppression of the WAL correction by magnetic field 
strongly depends on the direction of the field.
In particular, if the entire nodal line lies in one plane, as, for example, in $\text{ZrSiS}$ (see Fig.~\ref{Fig:Tubes}a),
the in-plane magnetic field can partially suppress the WAL correction, while the field perpendicular to the plane
has no significant effect.
We note also that for sufficiently weak disorder ($\mu\tau_{\text{tr}}\ggg1$) the conductivity is dominated
by the Drude contribution (\ref{eq:sigmaklow}), and the interference correction (\ref{WAL}) is negligible.

{\it Summary and discussion.}
We have studied the conductivity in a weakly disordered NLS with charged impurities.
At low temperature and chemical potential, the screened electrostatic interaction is qualitatively different from
that in conventional semiconductors and semimetals
and includes a broad regime of distances for which $\phi(r) \propto 1/r^2$ (see Fig.\ \ref{Fig:ScreeningPlot}).
Such screening of impurities can potentially be probed directly by
scanning tunnelling microscopy, and it also manifests itself in the dependencies of the conductivity on temperature
and doping level.

In particular, a low-doped NLS exhibits divergent quasiparticle mobility $\propto |\mu|^{-1}$ at vanishing chemical potential $\mu$, with conductivity
given by Eq.~(\ref{eq:sigmaklow}). In this regime, the NLS also exhibits strong temperature dependence of the conductivity, $\sigma(T) \propto T$, at $T \gg |\mu|$. At larger doping, the impurities are strongly screened, and the transport properties of the NLS
resemble those of a conventional metal.

In all of these regimes the conductivity of an NLS receives large WAL
corrections [Eq.~(\ref{WAL})]
due to the suppressed motion of the charge carriers along the nodal line, which makes electron dynamics effectively
2D on short scales and thus leads to strong single-particle interference effects.

For presently existing NLSs, which have $\ko$ of order $1$\,\AA$^{-1}$ and $v$ of order $10^8$\,cm/s, the energy scale $\alpha g v \ko$ is as large as several eV for $\alpha \sim 0.1$.  Existing NLS materials are therefore likely to fall in the ``low doping" regime of our analysis.  Recent experiments on ZrSiS\cite{Neupane:ZrSiS, Singha:ZrSiS, Ali:ZrSiS, FuhrerGroup:unpublished}, for example, report $\mu$ of order several hundred meV.  In this case Eq.\ (\ref{eq:sigmaklow}) implies a low-temperature resistivity on the order of tens of n$\Omega \cdot \textrm{cm}$, assuming a charged impurity concentration of order $10^{19}$\,cm$^{-3}$ (as reported in Ref.~\onlinecite{FuhrerGroup:unpublished}).  This is consistent with the experimental measurements in Refs.~\onlinecite{Neupane:ZrSiS, Ali:ZrSiS}. 

Finally, we note that current experiments correspond to the regime of weak disorder ($\mu\tau_0\gg1$),
which is assumed throughout this paper and which requires 
$n_{\text{imp}}\ll{g^2 \ko^2|\mu|}/{v}\sim 10^{21}\textrm{ cm}^{-3}$.
Since there is no localisation in systems of effectively-2D Dirac fermions
(in the absence of scattering between opposite ends of the nodal line), one can expect
that stronger disorder ($n_{\text{imp}}\gg {g^2 \ko^2|\mu|}/{v}$) may lead to a universal minimal
conductivity $\sigma\sim e^2\ko$ in an NLS. However, this regime of strong disorder is left for a future study.
Another question, which deserves further investigation, is the role of the interaction between quasiparticles in
transport, as for sufficiently high temperatures the NLS resistivity will be dominated by the
scattering of quasiparticles on each other rather than impurity scattering.

\

{\it Acknowledgements.}
We are grateful to M.S.~Fuhrer, B.~Weber and especially Ya.I.~Rodionov for valuable discussions.  
SVS was financially supported
by AFOSR, NSF QIS, ARO MURI, ARO, ARL CDQI and NSF PFC at Joint Quantum Institute. BS was supported as part of the MIT Center for Excitonics, an Energy Frontier Research Center funded by the U.S. Department of Energy, Office of Science, Basic Energy Sciences under Award no.\ DE-SC0001088.  SVS also acknowledges the hospitality of
School of Physics and Astronomy at Monash University and of the MIT Center for Excitonics, where parts of this work
were completed.


\bibliography{references}

\begin{thebibliography}{41}%
\makeatletter
\providecommand \@ifxundefined [1]{%
 \@ifx{#1\undefined}
}%
\providecommand \@ifnum [1]{%
 \ifnum #1\expandafter \@firstoftwo
 \else \expandafter \@secondoftwo
 \fi
}%
\providecommand \@ifx [1]{%
 \ifx #1\expandafter \@firstoftwo
 \else \expandafter \@secondoftwo
 \fi
}%
\providecommand \natexlab [1]{#1}%
\providecommand \enquote  [1]{``#1''}%
\providecommand \bibnamefont  [1]{#1}%
\providecommand \bibfnamefont [1]{#1}%
\providecommand \citenamefont [1]{#1}%
\providecommand \href@noop [0]{\@secondoftwo}%
\providecommand \href [0]{\begingroup \@sanitize@url \@href}%
\providecommand \@href[1]{\@@startlink{#1}\@@href}%
\providecommand \@@href[1]{\endgroup#1\@@endlink}%
\providecommand \@sanitize@url [0]{\catcode `\\12\catcode `\$12\catcode
  `\&12\catcode `\#12\catcode `\^12\catcode `\_12\catcode `\%12\relax}%
\providecommand \@@startlink[1]{}%
\providecommand \@@endlink[0]{}%
\providecommand \url  [0]{\begingroup\@sanitize@url \@url }%
\providecommand \@url [1]{\endgroup\@href {#1}{\urlprefix }}%
\providecommand \urlprefix  [0]{URL }%
\providecommand \Eprint [0]{\href }%
\providecommand \doibase [0]{http://dx.doi.org/}%
\providecommand \selectlanguage [0]{\@gobble}%
\providecommand \bibinfo  [0]{\@secondoftwo}%
\providecommand \bibfield  [0]{\@secondoftwo}%
\providecommand \translation [1]{[#1]}%
\providecommand \BibitemOpen [0]{}%
\providecommand \bibitemStop [0]{}%
\providecommand \bibitemNoStop [0]{.\EOS\space}%
\providecommand \EOS [0]{\spacefactor3000\relax}%
\providecommand \BibitemShut  [1]{\csname bibitem#1\endcsname}%
\let\auto@bib@innerbib\@empty
\bibitem [{\citenamefont {Wan}\ \emph {et~al.}(2011)\citenamefont {Wan},
  \citenamefont {Turner}, \citenamefont {Vishwanath},\ and\ \citenamefont
  {Savrasov}}]{Wan:WeylProp}%
  \BibitemOpen
  \bibfield  {author} {\bibinfo {author} {\bibfnamefont {Xiangang}\
  \bibnamefont {Wan}}, \bibinfo {author} {\bibfnamefont {Ari~M.}\ \bibnamefont
  {Turner}}, \bibinfo {author} {\bibfnamefont {Ashvin}\ \bibnamefont
  {Vishwanath}}, \ and\ \bibinfo {author} {\bibfnamefont {Sergey~Y.}\
  \bibnamefont {Savrasov}},\ }\bibfield  {title} {\enquote {\bibinfo {title}
  {Topological semimetal and {Fermi}-arc surface states in the electronic
  structure of pyrochlore iridates},}\ }\href@noop {} {\bibfield  {journal}
  {\bibinfo  {journal} {Phys. Rev. B}\ }\textbf {\bibinfo {volume} {83}},\
  \bibinfo {pages} {205101} (\bibinfo {year} {2011})}\BibitemShut {NoStop}%
\bibitem [{\citenamefont {Huang}\ \emph {et~al.}(2015)\citenamefont {Huang},
  \citenamefont {Xu}, \citenamefont {Belopolski}, \citenamefont {Lee},
  \citenamefont {Chang}, \citenamefont {Wang}, \citenamefont {Alidoust},
  \citenamefont {Bian}, \citenamefont {Neupane}, \citenamefont {Zhang},
  \citenamefont {Jia}, \citenamefont {Bansil}, \citenamefont {Lin},\ and\
  \citenamefont {Hasan}}]{ZHasan:TaAs}%
  \BibitemOpen
  \bibfield  {author} {\bibinfo {author} {\bibfnamefont {Shin-Ming}\
  \bibnamefont {Huang}}, \bibinfo {author} {\bibfnamefont {Su-Yang}\
  \bibnamefont {Xu}}, \bibinfo {author} {\bibfnamefont {Ilya}\ \bibnamefont
  {Belopolski}}, \bibinfo {author} {\bibfnamefont {Chi-Cheng}\ \bibnamefont
  {Lee}}, \bibinfo {author} {\bibfnamefont {Guoqing}\ \bibnamefont {Chang}},
  \bibinfo {author} {\bibfnamefont {BaoKai}\ \bibnamefont {Wang}}, \bibinfo
  {author} {\bibfnamefont {Nasser}\ \bibnamefont {Alidoust}}, \bibinfo {author}
  {\bibfnamefont {Guang}\ \bibnamefont {Bian}}, \bibinfo {author}
  {\bibfnamefont {Madhab}\ \bibnamefont {Neupane}}, \bibinfo {author}
  {\bibfnamefont {Chenglong}\ \bibnamefont {Zhang}}, \bibinfo {author}
  {\bibfnamefont {Shuang}\ \bibnamefont {Jia}}, \bibinfo {author}
  {\bibfnamefont {Arun}\ \bibnamefont {Bansil}}, \bibinfo {author}
  {\bibfnamefont {Hsin}\ \bibnamefont {Lin}}, \ and\ \bibinfo {author}
  {\bibfnamefont {M.~Zahid}\ \bibnamefont {Hasan}},\ }\bibfield  {title}
  {\enquote {\bibinfo {title} {A {Weyl} {Fermion} semimetal with surface
  {Fermi} arcs in the transition metal monopnictide {TaAs} class},}\
  }\href@noop {} {\bibfield  {journal} {\bibinfo  {journal} {Nature Comm.}\
  }\textbf {\bibinfo {volume} {6}},\ \bibinfo {pages} {7373} (\bibinfo {year}
  {2015})}\BibitemShut {NoStop}%
\bibitem [{\citenamefont {Xu}\ \emph {et~al.}(2015{\natexlab{a}})\citenamefont
  {Xu}, \citenamefont {Belopolski}, \citenamefont {Alidoust}, \citenamefont
  {Neupane}, \citenamefont {Bian}, \citenamefont {Zhang}, \citenamefont
  {Sankar}, \citenamefont {Chang}, \citenamefont {Yuan}, \citenamefont {Lee},
  \citenamefont {Huang}, \citenamefont {Zheng}, \citenamefont {Ma},
  \citenamefont {Sanchez}, \citenamefont {Wang}, \citenamefont {Bansil},
  \citenamefont {Chou}, \citenamefont {Shibayev}, \citenamefont {Lin},
  \citenamefont {Jia},\ and\ \citenamefont {Hasan}}]{ZHasan:TaAs2}%
  \BibitemOpen
  \bibfield  {author} {\bibinfo {author} {\bibfnamefont {Su-Yang}\ \bibnamefont
  {Xu}}, \bibinfo {author} {\bibfnamefont {Ilya}\ \bibnamefont {Belopolski}},
  \bibinfo {author} {\bibfnamefont {Nasser}\ \bibnamefont {Alidoust}}, \bibinfo
  {author} {\bibfnamefont {Madhab}\ \bibnamefont {Neupane}}, \bibinfo {author}
  {\bibfnamefont {Guang}\ \bibnamefont {Bian}}, \bibinfo {author}
  {\bibfnamefont {Chenglong}\ \bibnamefont {Zhang}}, \bibinfo {author}
  {\bibfnamefont {Raman}\ \bibnamefont {Sankar}}, \bibinfo {author}
  {\bibfnamefont {Guoqing}\ \bibnamefont {Chang}}, \bibinfo {author}
  {\bibfnamefont {Zhujun}\ \bibnamefont {Yuan}}, \bibinfo {author}
  {\bibfnamefont {Chi-Cheng}\ \bibnamefont {Lee}}, \bibinfo {author}
  {\bibfnamefont {Shin-Ming}\ \bibnamefont {Huang}}, \bibinfo {author}
  {\bibfnamefont {Hao}\ \bibnamefont {Zheng}}, \bibinfo {author} {\bibfnamefont
  {Jie}\ \bibnamefont {Ma}}, \bibinfo {author} {\bibfnamefont {Daniel~S.}\
  \bibnamefont {Sanchez}}, \bibinfo {author} {\bibfnamefont {BaoKai}\
  \bibnamefont {Wang}}, \bibinfo {author} {\bibfnamefont {Arun}\ \bibnamefont
  {Bansil}}, \bibinfo {author} {\bibfnamefont {Fangcheng}\ \bibnamefont
  {Chou}}, \bibinfo {author} {\bibfnamefont {Pavel~P.}\ \bibnamefont
  {Shibayev}}, \bibinfo {author} {\bibfnamefont {Hsin}\ \bibnamefont {Lin}},
  \bibinfo {author} {\bibfnamefont {Shuang}\ \bibnamefont {Jia}}, \ and\
  \bibinfo {author} {\bibfnamefont {M.~Zahid}\ \bibnamefont {Hasan}},\
  }\bibfield  {title} {\enquote {\bibinfo {title} {Discovery of a {Weyl}
  fermion semimetal and topological {Fermi} arcs},}\ }\href@noop {} {\bibfield
  {journal} {\bibinfo  {journal} {Science}\ }\textbf {\bibinfo {volume}
  {349}},\ \bibinfo {pages} {6248} (\bibinfo {year}
  {2015}{\natexlab{a}})}\BibitemShut {NoStop}%
\bibitem [{\citenamefont {Xu}\ \emph {et~al.}(2015{\natexlab{b}})\citenamefont
  {Xu}, \citenamefont {Belopolski}, \citenamefont {Sanchez}, \citenamefont
  {Zhang}, \citenamefont {Chang}, \citenamefont {Guo}, \citenamefont {Bian},
  \citenamefont {Yuan}, \citenamefont {Lu}, \citenamefont {Chang},
  \citenamefont {Shibayev}, \citenamefont {Prokopovych}, \citenamefont
  {Alidoust}, \citenamefont {Zheng}, \citenamefont {Lee}, \citenamefont
  {Huang}, \citenamefont {Sankar}, \citenamefont {Chou}, \citenamefont {Hsu},
  \citenamefont {Jeng}, \citenamefont {Bansil}, \citenamefont {Neupert},
  \citenamefont {Strocov}, \citenamefont {Lin}, \citenamefont {Jia},\ and\
  \citenamefont {Hasan}}]{ZHasan:TaP}%
  \BibitemOpen
  \bibfield  {author} {\bibinfo {author} {\bibfnamefont {Su-Yang}\ \bibnamefont
  {Xu}}, \bibinfo {author} {\bibfnamefont {Ilya}\ \bibnamefont {Belopolski}},
  \bibinfo {author} {\bibfnamefont {Daniel~S.}\ \bibnamefont {Sanchez}},
  \bibinfo {author} {\bibfnamefont {Chenglong}\ \bibnamefont {Zhang}}, \bibinfo
  {author} {\bibfnamefont {Guoqing}\ \bibnamefont {Chang}}, \bibinfo {author}
  {\bibfnamefont {Cheng}\ \bibnamefont {Guo}}, \bibinfo {author} {\bibfnamefont
  {Guang}\ \bibnamefont {Bian}}, \bibinfo {author} {\bibfnamefont {Zhujun}\
  \bibnamefont {Yuan}}, \bibinfo {author} {\bibfnamefont {Hong}\ \bibnamefont
  {Lu}}, \bibinfo {author} {\bibfnamefont {Tay-Rong}\ \bibnamefont {Chang}},
  \bibinfo {author} {\bibfnamefont {Pavel~P.}\ \bibnamefont {Shibayev}},
  \bibinfo {author} {\bibfnamefont {Mykhailo~L.}\ \bibnamefont {Prokopovych}},
  \bibinfo {author} {\bibfnamefont {Nasser}\ \bibnamefont {Alidoust}}, \bibinfo
  {author} {\bibfnamefont {Hao}\ \bibnamefont {Zheng}}, \bibinfo {author}
  {\bibfnamefont {Chi-Cheng}\ \bibnamefont {Lee}}, \bibinfo {author}
  {\bibfnamefont {Shin-Ming}\ \bibnamefont {Huang}}, \bibinfo {author}
  {\bibfnamefont {Raman}\ \bibnamefont {Sankar}}, \bibinfo {author}
  {\bibfnamefont {Fangcheng}\ \bibnamefont {Chou}}, \bibinfo {author}
  {\bibfnamefont {Chuang-Han}\ \bibnamefont {Hsu}}, \bibinfo {author}
  {\bibfnamefont {Horng-Tay}\ \bibnamefont {Jeng}}, \bibinfo {author}
  {\bibfnamefont {Arun}\ \bibnamefont {Bansil}}, \bibinfo {author}
  {\bibfnamefont {Titus}\ \bibnamefont {Neupert}}, \bibinfo {author}
  {\bibfnamefont {Vladimir~N.}\ \bibnamefont {Strocov}}, \bibinfo {author}
  {\bibfnamefont {Hsin}\ \bibnamefont {Lin}}, \bibinfo {author} {\bibfnamefont
  {Shuang}\ \bibnamefont {Jia}}, \ and\ \bibinfo {author} {\bibfnamefont
  {M.~Zahid}\ \bibnamefont {Hasan}},\ }\bibfield  {title} {\enquote {\bibinfo
  {title} {Experimental discovery of a topological {Weyl} semimetal state in
  {TaP}},}\ }\href@noop {} {\bibfield  {journal} {\bibinfo  {journal} {Sci.
  Adv.}\ }\textbf {\bibinfo {volume} {1}},\ \bibinfo {pages} {1501092}
  (\bibinfo {year} {2015}{\natexlab{b}})}\BibitemShut {NoStop}%
\bibitem [{\citenamefont {Xu}\ \emph {et~al.}(2015{\natexlab{c}})\citenamefont
  {Xu}, \citenamefont {Alidoust}, \citenamefont {Belopolski}, \citenamefont
  {Yuan}, \citenamefont {Bian}, \citenamefont {Chang}, \citenamefont {Zheng},
  \citenamefont {Strocov}, \citenamefont {Sanchez}, \citenamefont {Chang},
  \citenamefont {Zhang}, \citenamefont {Mou}, \citenamefont {Wu}, \citenamefont
  {Huang}, \citenamefont {Lee}, \citenamefont {Huang}, \citenamefont
  {BaoKaiWang}, \citenamefont {Bansil}, \citenamefont {Jeng}, \citenamefont
  {Neupert}, \citenamefont {Kaminski}, \citenamefont {Lin}, \citenamefont
  {Jia},\ and\ \citenamefont {Hasan}}]{ZHasan:NbAs}%
  \BibitemOpen
  \bibfield  {author} {\bibinfo {author} {\bibfnamefont {Su-Yang}\ \bibnamefont
  {Xu}}, \bibinfo {author} {\bibfnamefont {Nasser}\ \bibnamefont {Alidoust}},
  \bibinfo {author} {\bibfnamefont {Ilya}\ \bibnamefont {Belopolski}}, \bibinfo
  {author} {\bibfnamefont {Zhujun}\ \bibnamefont {Yuan}}, \bibinfo {author}
  {\bibfnamefont {Guang}\ \bibnamefont {Bian}}, \bibinfo {author}
  {\bibfnamefont {Tay-Rong}\ \bibnamefont {Chang}}, \bibinfo {author}
  {\bibfnamefont {Hao}\ \bibnamefont {Zheng}}, \bibinfo {author} {\bibfnamefont
  {Vladimir~N.}\ \bibnamefont {Strocov}}, \bibinfo {author} {\bibfnamefont
  {Daniel~S.}\ \bibnamefont {Sanchez}}, \bibinfo {author} {\bibfnamefont
  {Guoqing}\ \bibnamefont {Chang}}, \bibinfo {author} {\bibfnamefont
  {Chenglong}\ \bibnamefont {Zhang}}, \bibinfo {author} {\bibfnamefont
  {Daixiang}\ \bibnamefont {Mou}}, \bibinfo {author} {\bibfnamefont {Yun}\
  \bibnamefont {Wu}}, \bibinfo {author} {\bibfnamefont {Lunan}\ \bibnamefont
  {Huang}}, \bibinfo {author} {\bibfnamefont {Chi-Cheng}\ \bibnamefont {Lee}},
  \bibinfo {author} {\bibfnamefont {Shin-Ming}\ \bibnamefont {Huang}}, \bibinfo
  {author} {\bibnamefont {BaoKaiWang}}, \bibinfo {author} {\bibfnamefont
  {Arun}\ \bibnamefont {Bansil}}, \bibinfo {author} {\bibfnamefont {Horng-Tay}\
  \bibnamefont {Jeng}}, \bibinfo {author} {\bibfnamefont {Titus}\ \bibnamefont
  {Neupert}}, \bibinfo {author} {\bibfnamefont {Adam}\ \bibnamefont
  {Kaminski}}, \bibinfo {author} {\bibfnamefont {Hsin}\ \bibnamefont {Lin}},
  \bibinfo {author} {\bibfnamefont {Shuang}\ \bibnamefont {Jia}}, \ and\
  \bibinfo {author} {\bibfnamefont {M.~Zahid}\ \bibnamefont {Hasan}},\
  }\bibfield  {title} {\enquote {\bibinfo {title} {Discovery of a {Weyl}
  fermion state with {Fermi} arcs in niobium arsenide},}\ }\href@noop {}
  {\bibfield  {journal} {\bibinfo  {journal} {Nat. Phys.}\ }\textbf {\bibinfo
  {volume} {11}},\ \bibinfo {pages} {748} (\bibinfo {year}
  {2015}{\natexlab{c}})}\BibitemShut {NoStop}%
\bibitem [{\citenamefont {Lv}\ \emph {et~al.}(2015)\citenamefont {Lv},
  \citenamefont {Weng}, \citenamefont {Fu}, \citenamefont {Wang}, \citenamefont
  {Miao}, \citenamefont {Ma}, \citenamefont {Richard}, \citenamefont {Huang},
  \citenamefont {Zhao}, \citenamefont {Chen}, \citenamefont {Fang},
  \citenamefont {Dai}, \citenamefont {Qian},\ and\ \citenamefont
  {Ding}}]{Weng:PhotCrystWSM}%
  \BibitemOpen
  \bibfield  {author} {\bibinfo {author} {\bibfnamefont {B.~Q.}\ \bibnamefont
  {Lv}}, \bibinfo {author} {\bibfnamefont {H.~M.}\ \bibnamefont {Weng}},
  \bibinfo {author} {\bibfnamefont {B.~B.}\ \bibnamefont {Fu}}, \bibinfo
  {author} {\bibfnamefont {X.~P.}\ \bibnamefont {Wang}}, \bibinfo {author}
  {\bibfnamefont {H.}~\bibnamefont {Miao}}, \bibinfo {author} {\bibfnamefont
  {J.}~\bibnamefont {Ma}}, \bibinfo {author} {\bibfnamefont {P.}~\bibnamefont
  {Richard}}, \bibinfo {author} {\bibfnamefont {X.~C.}\ \bibnamefont {Huang}},
  \bibinfo {author} {\bibfnamefont {L.~X.}\ \bibnamefont {Zhao}}, \bibinfo
  {author} {\bibfnamefont {G.~F.}\ \bibnamefont {Chen}}, \bibinfo {author}
  {\bibfnamefont {Z.}~\bibnamefont {Fang}}, \bibinfo {author} {\bibfnamefont
  {X.}~\bibnamefont {Dai}}, \bibinfo {author} {\bibfnamefont {T.}~\bibnamefont
  {Qian}}, \ and\ \bibinfo {author} {\bibfnamefont {H.}~\bibnamefont {Ding}},\
  }\bibfield  {title} {\enquote {\bibinfo {title} {Experimental discovery of
  {Weyl} semimetal {TaAs}},}\ }\href@noop {} {\bibfield  {journal} {\bibinfo
  {journal} {Phys. Rev. X}\ }\textbf {\bibinfo {volume} {5}},\ \bibinfo {pages}
  {031013} (\bibinfo {year} {2015})}\BibitemShut {NoStop}%
\bibitem [{\citenamefont {Schoop}\ \emph {et~al.}(2016)\citenamefont {Schoop},
  \citenamefont {Ali}, \citenamefont {Stra{\ss}er}, \citenamefont {Topp},
  \citenamefont {Varykhalov}, \citenamefont {Marchenko}, \citenamefont
  {Duppel}, \citenamefont {Parkin}, \citenamefont {Lotsch},\ and\ \citenamefont
  {Ast}}]{ZrSiS:first}%
  \BibitemOpen
  \bibfield  {author} {\bibinfo {author} {\bibfnamefont {Leslie~M.}\
  \bibnamefont {Schoop}}, \bibinfo {author} {\bibfnamefont {Mazhar~N.}\
  \bibnamefont {Ali}}, \bibinfo {author} {\bibfnamefont {Carola}\ \bibnamefont
  {Stra{\ss}er}}, \bibinfo {author} {\bibfnamefont {Andreas}\ \bibnamefont
  {Topp}}, \bibinfo {author} {\bibfnamefont {Andrei}\ \bibnamefont
  {Varykhalov}}, \bibinfo {author} {\bibfnamefont {Dmitry}\ \bibnamefont
  {Marchenko}}, \bibinfo {author} {\bibfnamefont {Viola}\ \bibnamefont
  {Duppel}}, \bibinfo {author} {\bibfnamefont {Stuart S.~P.}\ \bibnamefont
  {Parkin}}, \bibinfo {author} {\bibfnamefont {Bettina~V.}\ \bibnamefont
  {Lotsch}}, \ and\ \bibinfo {author} {\bibfnamefont {Christian~R.}\
  \bibnamefont {Ast}},\ }\bibfield  {title} {\enquote {\bibinfo {title} {Dirac
  cone protected by non-symmorphic symmetry and three-dimensional {Dirac} line
  node in {ZrSiS}},}\ }\href@noop {} {\bibfield  {journal} {\bibinfo  {journal}
  {Nat. Comm.}\ }\textbf {\bibinfo {volume} {7}},\ \bibinfo {pages} {11696}
  (\bibinfo {year} {2016})}\BibitemShut {NoStop}%
\bibitem [{\citenamefont {Bian}\ \emph
  {et~al.}(2016{\natexlab{a}})\citenamefont {Bian}, \citenamefont {Chang},
  \citenamefont {Sankar}, \citenamefont {Xu}, \citenamefont {Zheng},
  \citenamefont {Neupert}, \citenamefont {Chiu}, \citenamefont {Huang},
  \citenamefont {Chang}, \citenamefont {Belopolski}, \citenamefont {Sanchez},
  \citenamefont {Neupane}, \citenamefont {Alidoust}, \citenamefont {Liu},
  \citenamefont {Wang}, \citenamefont {Lee}, \citenamefont {Jeng},
  \citenamefont {Zhang}, \citenamefont {Yuan}, \citenamefont {Jia},
  \citenamefont {Bansil}, \citenamefont {Chou}, \citenamefont {Lin},\ and\
  \citenamefont {Hasan}}]{PbTaSe2:first}%
  \BibitemOpen
  \bibfield  {author} {\bibinfo {author} {\bibfnamefont {Guang}\ \bibnamefont
  {Bian}}, \bibinfo {author} {\bibfnamefont {Tay-Rong}\ \bibnamefont {Chang}},
  \bibinfo {author} {\bibfnamefont {Raman}\ \bibnamefont {Sankar}}, \bibinfo
  {author} {\bibfnamefont {Su-Yang}\ \bibnamefont {Xu}}, \bibinfo {author}
  {\bibfnamefont {Hao}\ \bibnamefont {Zheng}}, \bibinfo {author} {\bibfnamefont
  {Titus}\ \bibnamefont {Neupert}}, \bibinfo {author} {\bibfnamefont
  {Ching-Kai}\ \bibnamefont {Chiu}}, \bibinfo {author} {\bibfnamefont
  {Shin-Ming}\ \bibnamefont {Huang}}, \bibinfo {author} {\bibfnamefont
  {Guoqing}\ \bibnamefont {Chang}}, \bibinfo {author} {\bibfnamefont {Ilya}\
  \bibnamefont {Belopolski}}, \bibinfo {author} {\bibfnamefont {Daniel~S.}\
  \bibnamefont {Sanchez}}, \bibinfo {author} {\bibfnamefont {Madhab}\
  \bibnamefont {Neupane}}, \bibinfo {author} {\bibfnamefont {Nasser}\
  \bibnamefont {Alidoust}}, \bibinfo {author} {\bibfnamefont {Chang}\
  \bibnamefont {Liu}}, \bibinfo {author} {\bibfnamefont {BaoKai}\ \bibnamefont
  {Wang}}, \bibinfo {author} {\bibfnamefont {Chi-Cheng}\ \bibnamefont {Lee}},
  \bibinfo {author} {\bibfnamefont {Horng-Tay}\ \bibnamefont {Jeng}}, \bibinfo
  {author} {\bibfnamefont {Chenglong}\ \bibnamefont {Zhang}}, \bibinfo {author}
  {\bibfnamefont {Zhujun}\ \bibnamefont {Yuan}}, \bibinfo {author}
  {\bibfnamefont {Shuang}\ \bibnamefont {Jia}}, \bibinfo {author}
  {\bibfnamefont {Arun}\ \bibnamefont {Bansil}}, \bibinfo {author}
  {\bibfnamefont {Fangcheng}\ \bibnamefont {Chou}}, \bibinfo {author}
  {\bibfnamefont {Hsin}\ \bibnamefont {Lin}}, \ and\ \bibinfo {author}
  {\bibfnamefont {M.~Zahid}\ \bibnamefont {Hasan}},\ }\bibfield  {title}
  {\enquote {\bibinfo {title} {Topological nodal-line fermions in spin-orbit
  metal {PbTaSe2}},}\ }\href@noop {} {\bibfield  {journal} {\bibinfo  {journal}
  {Nat. Comm.}\ }\textbf {\bibinfo {volume} {7}},\ \bibinfo {pages} {10556}
  (\bibinfo {year} {2016}{\natexlab{a}})}\BibitemShut {NoStop}%
\bibitem [{\citenamefont {Kondo}\ \emph {et~al.}(2015)\citenamefont {Kondo},
  \citenamefont {Nakayama}, \citenamefont {Chen}, \citenamefont {Ishikawa},
  \citenamefont {Moon}, \citenamefont {Yamamoto}, \citenamefont {Ota},
  \citenamefont {Malaeb}, \citenamefont {Kanai}, \citenamefont {Nakashima},
  \citenamefont {Ishida}, \citenamefont {Yoshida}, \citenamefont {Yamamoto},
  \citenamefont {Matsunami}, \citenamefont {Kimura}, \citenamefont {Inami},
  \citenamefont {Ono}, \citenamefont {Kumigashira}, \citenamefont {Nakatsuji},
  \citenamefont {Balents},\ and\ \citenamefont {Shin}}]{Shin:quadratic}%
  \BibitemOpen
  \bibfield  {author} {\bibinfo {author} {\bibfnamefont {Takeshi}\ \bibnamefont
  {Kondo}}, \bibinfo {author} {\bibfnamefont {M.}~\bibnamefont {Nakayama}},
  \bibinfo {author} {\bibfnamefont {R.}~\bibnamefont {Chen}}, \bibinfo {author}
  {\bibfnamefont {J.~J.}\ \bibnamefont {Ishikawa}}, \bibinfo {author}
  {\bibfnamefont {E.-G.}\ \bibnamefont {Moon}}, \bibinfo {author}
  {\bibfnamefont {T.}~\bibnamefont {Yamamoto}}, \bibinfo {author}
  {\bibfnamefont {Y.}~\bibnamefont {Ota}}, \bibinfo {author} {\bibfnamefont
  {W.}~\bibnamefont {Malaeb}}, \bibinfo {author} {\bibfnamefont
  {H.}~\bibnamefont {Kanai}}, \bibinfo {author} {\bibfnamefont
  {Y.}~\bibnamefont {Nakashima}}, \bibinfo {author} {\bibfnamefont
  {Y.}~\bibnamefont {Ishida}}, \bibinfo {author} {\bibfnamefont
  {R.}~\bibnamefont {Yoshida}}, \bibinfo {author} {\bibfnamefont
  {H.}~\bibnamefont {Yamamoto}}, \bibinfo {author} {\bibfnamefont
  {M.}~\bibnamefont {Matsunami}}, \bibinfo {author} {\bibfnamefont
  {S.}~\bibnamefont {Kimura}}, \bibinfo {author} {\bibfnamefont
  {N.}~\bibnamefont {Inami}}, \bibinfo {author} {\bibfnamefont
  {K.}~\bibnamefont {Ono}}, \bibinfo {author} {\bibfnamefont {H.}~\bibnamefont
  {Kumigashira}}, \bibinfo {author} {\bibfnamefont {S.}~\bibnamefont
  {Nakatsuji}}, \bibinfo {author} {\bibfnamefont {L.}~\bibnamefont {Balents}},
  \ and\ \bibinfo {author} {\bibfnamefont {S.}~\bibnamefont {Shin}},\
  }\bibfield  {title} {\enquote {\bibinfo {title} {Quadratic {Fermi} node in a
  {3D} strongly correlated semimetal},}\ }\href@noop {} {\bibfield  {journal}
  {\bibinfo  {journal} {Nature Communications}\ }\textbf {\bibinfo {volume}
  {6}},\ \bibinfo {pages} {10042} (\bibinfo {year} {2015})}\BibitemShut
  {NoStop}%
\bibitem [{\citenamefont {Parameswaran}\ \emph {et~al.}(2014)\citenamefont
  {Parameswaran}, \citenamefont {Grover}, \citenamefont {Abanin}, \citenamefont
  {Pesin},\ and\ \citenamefont {Vishwanath}}]{Parameswaran:Anomaly}%
  \BibitemOpen
  \bibfield  {author} {\bibinfo {author} {\bibfnamefont {S.~A.}\ \bibnamefont
  {Parameswaran}}, \bibinfo {author} {\bibfnamefont {T.}~\bibnamefont
  {Grover}}, \bibinfo {author} {\bibfnamefont {D.~A.}\ \bibnamefont {Abanin}},
  \bibinfo {author} {\bibfnamefont {D.~A.}\ \bibnamefont {Pesin}}, \ and\
  \bibinfo {author} {\bibfnamefont {A.}~\bibnamefont {Vishwanath}},\ }\bibfield
   {title} {\enquote {\bibinfo {title} {Probing the chiral anomaly with
  nonlocal transport in three-dimensional topological semimetals},}\
  }\href@noop {} {\bibfield  {journal} {\bibinfo  {journal} {Phys. Rev. X}\
  }\textbf {\bibinfo {volume} {4}},\ \bibinfo {pages} {031035} (\bibinfo {year}
  {2014})}\BibitemShut {NoStop}%
\bibitem [{\citenamefont {Fradkin}(1986)}]{Fradkin2}%
  \BibitemOpen
  \bibfield  {author} {\bibinfo {author} {\bibfnamefont {E.}~\bibnamefont
  {Fradkin}},\ }\bibfield  {title} {\enquote {\bibinfo {title} {Critical
  behavior of disordered degenerate semiconductors. i. models, symmetries, and
  formalism},}\ }\href@noop {} {\bibfield  {journal} {\bibinfo  {journal}
  {Phys. Rev. B}\ }\textbf {\bibinfo {volume} {33}},\ \bibinfo {pages} {3257}
  (\bibinfo {year} {1986})}\BibitemShut {NoStop}%
\bibitem [{Note1()}]{Note1}%
  \BibitemOpen
  \bibinfo {note} {See Ref.~\protect \rev@citealpnum {Syzranov:review} for a
  review}\BibitemShut {NoStop}%
\bibitem [{\citenamefont {Takane}\ \emph {et~al.}(2016)\citenamefont {Takane},
  \citenamefont {Wang}, \citenamefont {Souma}, \citenamefont {Nakayama},
  \citenamefont {Trang}, \citenamefont {Sato}, \citenamefont {Takahashi},\ and\
  \citenamefont {Ando}}]{Ando:HfSiS}%
  \BibitemOpen
  \bibfield  {author} {\bibinfo {author} {\bibfnamefont {D.}~\bibnamefont
  {Takane}}, \bibinfo {author} {\bibfnamefont {Zhiwei}\ \bibnamefont {Wang}},
  \bibinfo {author} {\bibfnamefont {S.}~\bibnamefont {Souma}}, \bibinfo
  {author} {\bibfnamefont {K.}~\bibnamefont {Nakayama}}, \bibinfo {author}
  {\bibfnamefont {C.~X.}\ \bibnamefont {Trang}}, \bibinfo {author}
  {\bibfnamefont {T.}~\bibnamefont {Sato}}, \bibinfo {author} {\bibfnamefont
  {T.}~\bibnamefont {Takahashi}}, \ and\ \bibinfo {author} {\bibfnamefont
  {Yoichi}\ \bibnamefont {Ando}},\ }\bibfield  {title} {\enquote {\bibinfo
  {title} {Dirac-node arc in the topological line-node semimetal
  {Hf}{Si}{S}},}\ }\href@noop {} {\bibfield  {journal} {\bibinfo  {journal}
  {Phys. Rev. B}\ }\textbf {\bibinfo {volume} {94}},\ \bibinfo {pages} {121108}
  (\bibinfo {year} {2016})}\BibitemShut {NoStop}%
\bibitem [{\citenamefont {Burkov}\ \emph {et~al.}(2011)\citenamefont {Burkov},
  \citenamefont {Hook},\ and\ \citenamefont {Balents}}]{BurkovHookBalents}%
  \BibitemOpen
  \bibfield  {author} {\bibinfo {author} {\bibfnamefont {A.~A.}\ \bibnamefont
  {Burkov}}, \bibinfo {author} {\bibfnamefont {M.~D.}\ \bibnamefont {Hook}}, \
  and\ \bibinfo {author} {\bibfnamefont {Leon}\ \bibnamefont {Balents}},\
  }\bibfield  {title} {\enquote {\bibinfo {title} {Topological nodal
  semimetals},}\ }\href@noop {} {\bibfield  {journal} {\bibinfo  {journal}
  {Phys. Rev. B}\ }\textbf {\bibinfo {volume} {84}},\ \bibinfo {pages} {235126}
  (\bibinfo {year} {2011})}\BibitemShut {NoStop}%
\bibitem [{\citenamefont {Kim}\ \emph {et~al.}(2015)\citenamefont {Kim},
  \citenamefont {Wieder}, \citenamefont {Kane},\ and\ \citenamefont
  {Rappe}}]{KimKane:prediction}%
  \BibitemOpen
  \bibfield  {author} {\bibinfo {author} {\bibfnamefont {Youngkuk}\
  \bibnamefont {Kim}}, \bibinfo {author} {\bibfnamefont {Benjamin~J.}\
  \bibnamefont {Wieder}}, \bibinfo {author} {\bibfnamefont {C.~L.}\
  \bibnamefont {Kane}}, \ and\ \bibinfo {author} {\bibfnamefont {Andrew~M.}\
  \bibnamefont {Rappe}},\ }\bibfield  {title} {\enquote {\bibinfo {title}
  {Dirac line nodes in inversion-symmetric crystals},}\ }\href@noop {}
  {\bibfield  {journal} {\bibinfo  {journal} {Phys. Rev. Lett.}\ }\textbf
  {\bibinfo {volume} {115}},\ \bibinfo {pages} {036806} (\bibinfo {year}
  {2015})}\BibitemShut {NoStop}%
\bibitem [{\citenamefont {Xie}\ \emph {et~al.}(2015)\citenamefont {Xie},
  \citenamefont {Schoop}, \citenamefont {Seibel}, \citenamefont {Gibson},
  \citenamefont {Xie},\ and\ \citenamefont {Cava}}]{Xie:Ca3P2prediction}%
  \BibitemOpen
  \bibfield  {author} {\bibinfo {author} {\bibfnamefont {Lilia~S.}\
  \bibnamefont {Xie}}, \bibinfo {author} {\bibfnamefont {Leslie~M.}\
  \bibnamefont {Schoop}}, \bibinfo {author} {\bibfnamefont {Elizabeth~M.}\
  \bibnamefont {Seibel}}, \bibinfo {author} {\bibfnamefont {Quinn~D.}\
  \bibnamefont {Gibson}}, \bibinfo {author} {\bibfnamefont {Weiwei}\
  \bibnamefont {Xie}}, \ and\ \bibinfo {author} {\bibfnamefont {Robert~J.}\
  \bibnamefont {Cava}},\ }\bibfield  {title} {\enquote {\bibinfo {title} {A new
  form of {Ca3P2} with a ring of {Dirac} nodes},}\ }\href@noop {} {\bibfield
  {journal} {\bibinfo  {journal} {APL Mat.}\ }\textbf {\bibinfo {volume} {3}},\
  \bibinfo {pages} {083602} (\bibinfo {year} {2015})}\BibitemShut {NoStop}%
\bibitem [{\citenamefont {Fang}\ \emph {et~al.}(2015)\citenamefont {Fang},
  \citenamefont {Chen}, \citenamefont {Kee},\ and\ \citenamefont
  {Fu}}]{FangFu:prediction}%
  \BibitemOpen
  \bibfield  {author} {\bibinfo {author} {\bibfnamefont {Chen}\ \bibnamefont
  {Fang}}, \bibinfo {author} {\bibfnamefont {Yige}\ \bibnamefont {Chen}},
  \bibinfo {author} {\bibfnamefont {Hae-Young}\ \bibnamefont {Kee}}, \ and\
  \bibinfo {author} {\bibfnamefont {Liang}\ \bibnamefont {Fu}},\ }\bibfield
  {title} {\enquote {\bibinfo {title} {Topological nodal line semimetals with
  and without spin-orbital coupling},}\ }\href@noop {} {\bibfield  {journal}
  {\bibinfo  {journal} {Phys. Rev. B}\ }\textbf {\bibinfo {volume} {92}},\
  \bibinfo {pages} {081201(R)} (\bibinfo {year} {2015})}\BibitemShut {NoStop}%
\bibitem [{\citenamefont {Gan}\ \emph {et~al.}(2016)\citenamefont {Gan},
  \citenamefont {Wang}, \citenamefont {Jin}, \citenamefont {Ling},
  \citenamefont {Zhao}, \citenamefont {Xu}, \citenamefont {Liu},\ and\
  \citenamefont {Xu}}]{Gan:XB6}%
  \BibitemOpen
  \bibfield  {author} {\bibinfo {author} {\bibfnamefont {L.-Y.}\ \bibnamefont
  {Gan}}, \bibinfo {author} {\bibfnamefont {R.}~\bibnamefont {Wang}}, \bibinfo
  {author} {\bibfnamefont {Y.~J.}\ \bibnamefont {Jin}}, \bibinfo {author}
  {\bibfnamefont {D.~B.}\ \bibnamefont {Ling}}, \bibinfo {author}
  {\bibfnamefont {J.~Z.}\ \bibnamefont {Zhao}}, \bibinfo {author}
  {\bibfnamefont {W.~P.}\ \bibnamefont {Xu}}, \bibinfo {author} {\bibfnamefont
  {J.~F.}\ \bibnamefont {Liu}}, \ and\ \bibinfo {author} {\bibfnamefont
  {H.}~\bibnamefont {Xu}},\ }\href@noop {} {} (\bibinfo {year} {2016}),\
  \bibinfo {note} {arXiv:1611.06386}\BibitemShut {NoStop}%
\bibitem [{\citenamefont {Yu}\ \emph {et~al.}(2015)\citenamefont {Yu},
  \citenamefont {Weng}, \citenamefont {Fang}, \citenamefont {Dai},\ and\
  \citenamefont {Hu}}]{Yu:Cu3PdN}%
  \BibitemOpen
  \bibfield  {author} {\bibinfo {author} {\bibfnamefont {Rui}\ \bibnamefont
  {Yu}}, \bibinfo {author} {\bibfnamefont {Hongming}\ \bibnamefont {Weng}},
  \bibinfo {author} {\bibfnamefont {Zhong}\ \bibnamefont {Fang}}, \bibinfo
  {author} {\bibfnamefont {Xi}~\bibnamefont {Dai}}, \ and\ \bibinfo {author}
  {\bibfnamefont {Xiao}\ \bibnamefont {Hu}},\ }\bibfield  {title} {\enquote
  {\bibinfo {title} {Topological node-line semimetal and dirac semimetal state
  in antiperovskite {${\mathrm{Cu}}_{3}\mathrm{PdN}$}},}\ }\href@noop {}
  {\bibfield  {journal} {\bibinfo  {journal} {Phys. Rev. Lett.}\ }\textbf
  {\bibinfo {volume} {115}},\ \bibinfo {pages} {036807} (\bibinfo {year}
  {2015})}\BibitemShut {NoStop}%
\bibitem [{\citenamefont {Mullen}\ \emph {et~al.}(2015)\citenamefont {Mullen},
  \citenamefont {Uchoa},\ and\ \citenamefont
  {Glatzhofer}}]{Mullen:hyperhoneycomb}%
  \BibitemOpen
  \bibfield  {author} {\bibinfo {author} {\bibfnamefont {Kieran}\ \bibnamefont
  {Mullen}}, \bibinfo {author} {\bibfnamefont {Bruno}\ \bibnamefont {Uchoa}}, \
  and\ \bibinfo {author} {\bibfnamefont {Daniel~T.}\ \bibnamefont
  {Glatzhofer}},\ }\bibfield  {title} {\enquote {\bibinfo {title} {Line of
  {Dirac} nodes in hyperhoneycomb lattices},}\ }\href@noop {} {\bibfield
  {journal} {\bibinfo  {journal} {Phys. Rev. Lett.}\ }\textbf {\bibinfo
  {volume} {115}},\ \bibinfo {pages} {026403} (\bibinfo {year}
  {2015})}\BibitemShut {NoStop}%
\bibitem [{\citenamefont {Bian}\ \emph
  {et~al.}(2016{\natexlab{b}})\citenamefont {Bian}, \citenamefont {Chang},
  \citenamefont {Zheng}, \citenamefont {Velury}, \citenamefont {Xu},
  \citenamefont {Neupert}, \citenamefont {Chiu}, \citenamefont {Huang},
  \citenamefont {Sanchez}, \citenamefont {Belopolski}, \citenamefont
  {Alidoust}, \citenamefont {Chen}, \citenamefont {Chang}, \citenamefont
  {Bansil}, \citenamefont {Jeng}, \citenamefont {Lin},\ and\ \citenamefont
  {Hasan}}]{ZHasan:TlTaSe2}%
  \BibitemOpen
  \bibfield  {author} {\bibinfo {author} {\bibfnamefont {Guang}\ \bibnamefont
  {Bian}}, \bibinfo {author} {\bibfnamefont {Tay-Rong}\ \bibnamefont {Chang}},
  \bibinfo {author} {\bibfnamefont {Hao}\ \bibnamefont {Zheng}}, \bibinfo
  {author} {\bibfnamefont {Saavanth}\ \bibnamefont {Velury}}, \bibinfo {author}
  {\bibfnamefont {Su-Yang}\ \bibnamefont {Xu}}, \bibinfo {author}
  {\bibfnamefont {Titus}\ \bibnamefont {Neupert}}, \bibinfo {author}
  {\bibfnamefont {Ching-Kai}\ \bibnamefont {Chiu}}, \bibinfo {author}
  {\bibfnamefont {Shin-Ming}\ \bibnamefont {Huang}}, \bibinfo {author}
  {\bibfnamefont {Daniel~S.}\ \bibnamefont {Sanchez}}, \bibinfo {author}
  {\bibfnamefont {Ilya}\ \bibnamefont {Belopolski}}, \bibinfo {author}
  {\bibfnamefont {Nasser}\ \bibnamefont {Alidoust}}, \bibinfo {author}
  {\bibfnamefont {Peng-Jen}\ \bibnamefont {Chen}}, \bibinfo {author}
  {\bibfnamefont {Guoqing}\ \bibnamefont {Chang}}, \bibinfo {author}
  {\bibfnamefont {Arun}\ \bibnamefont {Bansil}}, \bibinfo {author}
  {\bibfnamefont {Horng-Tay}\ \bibnamefont {Jeng}}, \bibinfo {author}
  {\bibfnamefont {Hsin}\ \bibnamefont {Lin}}, \ and\ \bibinfo {author}
  {\bibfnamefont {M.~Zahid}\ \bibnamefont {Hasan}},\ }\bibfield  {title}
  {\enquote {\bibinfo {title} {Drumhead surface states and topological
  nodal-line fermions in {${\mathrm{TlTaSe}}_{2}$}},}\ }\href@noop {}
  {\bibfield  {journal} {\bibinfo  {journal} {Phys. Rev. B}\ }\textbf {\bibinfo
  {volume} {93}},\ \bibinfo {pages} {121113} (\bibinfo {year}
  {2016}{\natexlab{b}})}\BibitemShut {NoStop}%
\bibitem [{\citenamefont {Skinner}(2014)}]{Skinner:WeylImp}%
  \BibitemOpen
  \bibfield  {author} {\bibinfo {author} {\bibfnamefont {Brian}\ \bibnamefont
  {Skinner}},\ }\bibfield  {title} {\enquote {\bibinfo {title} {Coulomb
  disorder in three-dimensional {Dirac} systems},}\ }\href@noop {} {\bibfield
  {journal} {\bibinfo  {journal} {Phys. Rev. B}\ }\textbf {\bibinfo {volume}
  {90}},\ \bibinfo {pages} {060202(R)} (\bibinfo {year} {2014})}\BibitemShut
  {NoStop}%
\bibitem [{\citenamefont {Rodionov}\ and\ \citenamefont
  {Syzranov}(2015)}]{Rodionov:donorsacceptors}%
  \BibitemOpen
  \bibfield  {author} {\bibinfo {author} {\bibfnamefont {Ya.~I.}\ \bibnamefont
  {Rodionov}}\ and\ \bibinfo {author} {\bibfnamefont {S.~V.}\ \bibnamefont
  {Syzranov}},\ }\bibfield  {title} {\enquote {\bibinfo {title} {Conductivity
  of a {Weyl} semimetal with donor and acceptor impurities},}\ }\href@noop {}
  {\bibfield  {journal} {\bibinfo  {journal} {Phys. Rev. B}\ }\textbf {\bibinfo
  {volume} {91}},\ \bibinfo {pages} {195107} (\bibinfo {year}
  {2015})}\BibitemShut {NoStop}%
\bibitem [{\citenamefont {Ando}(2006)}]{Ando:polarisation}%
  \BibitemOpen
  \bibfield  {author} {\bibinfo {author} {\bibfnamefont {Tsuneya}\ \bibnamefont
  {Ando}},\ }\bibfield  {title} {\enquote {\bibinfo {title} {Screening effect
  and impurity scattering in monolayer graphene},}\ }\href@noop {} {\bibfield
  {journal} {\bibinfo  {journal} {J. Phys. Soc. Japan}\ }\textbf {\bibinfo
  {volume} {75}},\ \bibinfo {pages} {074716} (\bibinfo {year}
  {2006})}\BibitemShut {NoStop}%
\bibitem [{\citenamefont {Hwang}\ and\ \citenamefont
  {Sarma}(2007)}]{HwangSarma:GraphenePolarisation}%
  \BibitemOpen
  \bibfield  {author} {\bibinfo {author} {\bibfnamefont {E.~H.}\ \bibnamefont
  {Hwang}}\ and\ \bibinfo {author} {\bibfnamefont {S.~Das}\ \bibnamefont
  {Sarma}},\ }\bibfield  {title} {\enquote {\bibinfo {title} {Dielectric
  function, screening, and plasmons in two-dimensional graphene},}\ }\href@noop
  {} {\bibfield  {journal} {\bibinfo  {journal} {Phys. Rev. B}\ }\textbf
  {\bibinfo {volume} {75}},\ \bibinfo {pages} {205418} (\bibinfo {year}
  {2007})}\BibitemShut {NoStop}%
\bibitem [{Note2()}]{Note2}%
  \BibitemOpen
  \bibinfo {note} {These constants of proportionality between $\Pi ({\protect
  \bf q})$ and $|{\protect \bf q}|$ are calculated for the case of a circular
  nodal line in Ref.~\protect \rev@citealpnum {Huh:NLS}.}\BibitemShut {Stop}%
\bibitem [{\citenamefont {Abrikosov}(1988)}]{Abrikosov:metals}%
  \BibitemOpen
  \bibfield  {author} {\bibinfo {author} {\bibfnamefont {A.~A.}\ \bibnamefont
  {Abrikosov}},\ }\href@noop {} {\emph {\bibinfo {title} {Fundamentals of the
  Theory of Metals}}}\ (\bibinfo  {publisher} {Elsevier},\ \bibinfo {address}
  {Oxford},\ \bibinfo {year} {1988})\BibitemShut {NoStop}%
\bibitem [{\citenamefont {Efetov}(1999)}]{Efetov:book}%
  \BibitemOpen
  \bibfield  {author} {\bibinfo {author} {\bibfnamefont {K.~B.}\ \bibnamefont
  {Efetov}},\ }\href@noop {} {\emph {\bibinfo {title} {Supersymetry in Disorder
  and Chaos}}}\ (\bibinfo  {publisher} {Cambridge University Press},\ \bibinfo
  {address} {New York},\ \bibinfo {year} {1999})\BibitemShut {NoStop}%
\bibitem [{\citenamefont {Abrikosov}\ \emph {et~al.}(1975)\citenamefont
  {Abrikosov}, \citenamefont {Gorkov},\ and\ \citenamefont
  {Dzyaloshinski}}]{AGD}%
  \BibitemOpen
  \bibfield  {author} {\bibinfo {author} {\bibfnamefont {A.~A.}\ \bibnamefont
  {Abrikosov}}, \bibinfo {author} {\bibfnamefont {L.~P.}\ \bibnamefont
  {Gorkov}}, \ and\ \bibinfo {author} {\bibfnamefont {I.~E.}\ \bibnamefont
  {Dzyaloshinski}},\ }\href@noop {} {\emph {\bibinfo {title} {Methods of
  Quantum Field Theory in Statistical Physics}}}\ (\bibinfo  {publisher}
  {Dover},\ \bibinfo {address} {New York},\ \bibinfo {year} {1975})\BibitemShut
  {NoStop}%
\bibitem [{Note3()}]{Note3}%
  \BibitemOpen
  \bibinfo {note} {This transport scattering is equivalent to the scattering of
  electrons on the surface of a 3D topological insulator with bulk
  impurities.\cite {Skinner:TIsurface}.}\BibitemShut {Stop}%
\bibitem [{\citenamefont {Gantmakher}(2005)}]{Gantmakher:book}%
  \BibitemOpen
  \bibfield  {author} {\bibinfo {author} {\bibfnamefont {V.~F.}\ \bibnamefont
  {Gantmakher}},\ }\href@noop {} {\emph {\bibinfo {title} {Electrons and
  Disorder in Solids}}}\ (\bibinfo  {publisher} {Oxford University Press},\
  \bibinfo {year} {2005})\BibitemShut {NoStop}%
\bibitem [{\citenamefont {McCann}\ \emph {et~al.}(2006)\citenamefont {McCann},
  \citenamefont {Kechedzhi}, \citenamefont {{Fal'ko}}, \citenamefont {Suzuura},
  \citenamefont {Ando},\ and\ \citenamefont {Altshuler}}]{McCann:WL}%
  \BibitemOpen
  \bibfield  {author} {\bibinfo {author} {\bibfnamefont {E.}~\bibnamefont
  {McCann}}, \bibinfo {author} {\bibfnamefont {K.}~\bibnamefont {Kechedzhi}},
  \bibinfo {author} {\bibfnamefont {Vladimir~I.}\ \bibnamefont {{Fal'ko}}},
  \bibinfo {author} {\bibfnamefont {H.}~\bibnamefont {Suzuura}}, \bibinfo
  {author} {\bibfnamefont {T.}~\bibnamefont {Ando}}, \ and\ \bibinfo {author}
  {\bibfnamefont {B.~L.}\ \bibnamefont {Altshuler}},\ }\bibfield  {title}
  {\enquote {\bibinfo {title} {Weak-localization magnetoresistance and valley
  symmetry in graphene},}\ }\href@noop {} {\bibfield  {journal} {\bibinfo
  {journal} {Phys. Rev. Lett.}\ }\textbf {\bibinfo {volume} {97}},\ \bibinfo
  {pages} {146805} (\bibinfo {year} {2006})}\BibitemShut {NoStop}%
\bibitem [{\citenamefont {Aleiner}\ and\ \citenamefont
  {Efetov}(2006)}]{AleinerEfetov}%
  \BibitemOpen
  \bibfield  {author} {\bibinfo {author} {\bibfnamefont {I.~L.}\ \bibnamefont
  {Aleiner}}\ and\ \bibinfo {author} {\bibfnamefont {K.~B.}\ \bibnamefont
  {Efetov}},\ }\bibfield  {title} {\enquote {\bibinfo {title} {A
  finite-temperature phase transition for disordered weakly interacting bosons
  in one dimension},}\ }\href@noop {} {\bibfield  {journal} {\bibinfo
  {journal} {Phys. Rev. Lett.}\ }\textbf {\bibinfo {volume} {97}},\ \bibinfo
  {pages} {236801} (\bibinfo {year} {2006})}\BibitemShut {NoStop}%
\bibitem [{\citenamefont {Neupane}\ \emph {et~al.}(2016)\citenamefont
  {Neupane}, \citenamefont {Belopolski}, \citenamefont {Hosen}, \citenamefont
  {Sanchez}, \citenamefont {Sankar}, \citenamefont {Szlawska}, \citenamefont
  {Xu}, \citenamefont {Dimitri}, \citenamefont {Dhakal}, \citenamefont
  {Maldonado}, \citenamefont {Oppeneer}, \citenamefont {Kaczorowski},
  \citenamefont {Chou}, \citenamefont {Hasan},\ and\ \citenamefont
  {Durakiewicz}}]{Neupane:ZrSiS}%
  \BibitemOpen
  \bibfield  {author} {\bibinfo {author} {\bibfnamefont {Madhab}\ \bibnamefont
  {Neupane}}, \bibinfo {author} {\bibfnamefont {Ilya}\ \bibnamefont
  {Belopolski}}, \bibinfo {author} {\bibfnamefont {M.~Mofazzel}\ \bibnamefont
  {Hosen}}, \bibinfo {author} {\bibfnamefont {Daniel~S.}\ \bibnamefont
  {Sanchez}}, \bibinfo {author} {\bibfnamefont {Raman}\ \bibnamefont {Sankar}},
  \bibinfo {author} {\bibfnamefont {Maria}\ \bibnamefont {Szlawska}}, \bibinfo
  {author} {\bibfnamefont {Su-Yang}\ \bibnamefont {Xu}}, \bibinfo {author}
  {\bibfnamefont {Klauss}\ \bibnamefont {Dimitri}}, \bibinfo {author}
  {\bibfnamefont {Nagendra}\ \bibnamefont {Dhakal}}, \bibinfo {author}
  {\bibfnamefont {Pablo}\ \bibnamefont {Maldonado}}, \bibinfo {author}
  {\bibfnamefont {Peter~M.}\ \bibnamefont {Oppeneer}}, \bibinfo {author}
  {\bibfnamefont {Dariusz}\ \bibnamefont {Kaczorowski}}, \bibinfo {author}
  {\bibfnamefont {Fangcheng}\ \bibnamefont {Chou}}, \bibinfo {author}
  {\bibfnamefont {M.~Zahid}\ \bibnamefont {Hasan}}, \ and\ \bibinfo {author}
  {\bibfnamefont {Tomasz}\ \bibnamefont {Durakiewicz}},\ }\bibfield  {title}
  {\enquote {\bibinfo {title} {Observation of topological nodal fermion
  semimetal phase in {ZrSiS}},}\ }\href@noop {} {\bibfield  {journal} {\bibinfo
   {journal} {Phys. Rev. B}\ }\textbf {\bibinfo {volume} {93}},\ \bibinfo
  {pages} {201104} (\bibinfo {year} {2016})}\BibitemShut {NoStop}%
\bibitem [{\citenamefont {Singha}\ \emph {et~al.}(2016)\citenamefont {Singha},
  \citenamefont {Pariari}, \citenamefont {Satpati},\ and\ \citenamefont
  {Mandal}}]{Singha:ZrSiS}%
  \BibitemOpen
  \bibfield  {author} {\bibinfo {author} {\bibfnamefont {R.}~\bibnamefont
  {Singha}}, \bibinfo {author} {\bibfnamefont {A.}~\bibnamefont {Pariari}},
  \bibinfo {author} {\bibfnamefont {B.}~\bibnamefont {Satpati}}, \ and\
  \bibinfo {author} {\bibfnamefont {P.}~\bibnamefont {Mandal}},\ }\href@noop {}
  {\enquote {\bibinfo {title} {Titanic magnetoresistance and signature of
  non-degenerate {D}irac nodes in {Zr}{Si}{S}},}\ } (\bibinfo {year} {2016}),\
  \bibinfo {note} {arxiv:1602.01993}\BibitemShut {NoStop}%
\bibitem [{\citenamefont {Ali}\ \emph {et~al.}(2016)\citenamefont {Ali},
  \citenamefont {Schoop}, \citenamefont {Garg}, \citenamefont {Lippmann},
  \citenamefont {Lara}, \citenamefont {Lotsch},\ and\ \citenamefont
  {Parkin}}]{Ali:ZrSiS}%
  \BibitemOpen
  \bibfield  {author} {\bibinfo {author} {\bibfnamefont {Mazhar~N.}\
  \bibnamefont {Ali}}, \bibinfo {author} {\bibfnamefont {Leslie~M.}\
  \bibnamefont {Schoop}}, \bibinfo {author} {\bibfnamefont {Chirag}\
  \bibnamefont {Garg}}, \bibinfo {author} {\bibfnamefont {Judith~M.}\
  \bibnamefont {Lippmann}}, \bibinfo {author} {\bibfnamefont {Eric}\
  \bibnamefont {Lara}}, \bibinfo {author} {\bibfnamefont {Bettina}\
  \bibnamefont {Lotsch}}, \ and\ \bibinfo {author} {\bibfnamefont {Stuart}\
  \bibnamefont {Parkin}},\ }\href@noop {} {\enquote {\bibinfo {title}
  {Butterfly magnetoresistance, quasi-{2D} {D}irac {F}ermi surfaces, and a
  topological phase transition in {Zr}{Si}{S}},}\ } (\bibinfo {year} {2016}),\
  \bibinfo {note} {arxiv:1603.09318}\BibitemShut {NoStop}%
\bibitem [{\citenamefont {Lodge}\ \emph {et~al.}(2017)\citenamefont {Lodge},
  \citenamefont {Chang}, \citenamefont {Singh}, \citenamefont {Hellerstedt},
  \citenamefont {Edmonds}, \citenamefont {Kaczorowski}, \citenamefont {Hosen},
  \citenamefont {Neupane}, \citenamefont {Lin}, \citenamefont {Fuhrer},
  \citenamefont {Weber},\ and\ \citenamefont
  {Ishigami}}]{FuhrerGroup:unpublished}%
  \BibitemOpen
  \bibfield  {author} {\bibinfo {author} {\bibfnamefont {Michael~S.}\
  \bibnamefont {Lodge}}, \bibinfo {author} {\bibfnamefont {Guoqing}\
  \bibnamefont {Chang}}, \bibinfo {author} {\bibfnamefont {Bahadur}\
  \bibnamefont {Singh}}, \bibinfo {author} {\bibfnamefont {Jack}\ \bibnamefont
  {Hellerstedt}}, \bibinfo {author} {\bibfnamefont {Mark}\ \bibnamefont
  {Edmonds}}, \bibinfo {author} {\bibfnamefont {Dariusz}\ \bibnamefont
  {Kaczorowski}}, \bibinfo {author} {\bibfnamefont {Md~Mofazzel}\ \bibnamefont
  {Hosen}}, \bibinfo {author} {\bibfnamefont {Madhab}\ \bibnamefont {Neupane}},
  \bibinfo {author} {\bibfnamefont {Hsin}\ \bibnamefont {Lin}}, \bibinfo
  {author} {\bibfnamefont {Michael~S.}\ \bibnamefont {Fuhrer}}, \bibinfo
  {author} {\bibfnamefont {Bent}\ \bibnamefont {Weber}}, \ and\ \bibinfo
  {author} {\bibfnamefont {Masahiro}\ \bibnamefont {Ishigami}},\ }\href@noop {}
  {} (\bibinfo {year} {2017}),\ \bibinfo {note} {coming soon}\BibitemShut
  {NoStop}%
\bibitem [{\citenamefont {Syzranov}\ and\ \citenamefont
  {Radzihovsky}(2016)}]{Syzranov:review}%
  \BibitemOpen
  \bibfield  {author} {\bibinfo {author} {\bibfnamefont {S.~V.}\ \bibnamefont
  {Syzranov}}\ and\ \bibinfo {author} {\bibfnamefont {L.}~\bibnamefont
  {Radzihovsky}},\ }\href@noop {} {\enquote {\bibinfo {title} {High-dimensional
  disorder-driven phenomena in {Weyl} semimetals, semiconductors and related
  systems},}\ } (\bibinfo {year} {2016}),\ \bibinfo {note}
  {arXiv:1609.05694}\BibitemShut {NoStop}%
\bibitem [{\citenamefont {Huh}\ \emph {et~al.}(2016)\citenamefont {Huh},
  \citenamefont {Moon},\ and\ \citenamefont {Kim}}]{Huh:NLS}%
  \BibitemOpen
  \bibfield  {author} {\bibinfo {author} {\bibfnamefont {Yejin}\ \bibnamefont
  {Huh}}, \bibinfo {author} {\bibfnamefont {Eun-Gook}\ \bibnamefont {Moon}}, \
  and\ \bibinfo {author} {\bibfnamefont {Yong~Baek}\ \bibnamefont {Kim}},\
  }\bibfield  {title} {\enquote {\bibinfo {title} {Long-range {Coulomb}
  interaction in nodal-ring semimetals},}\ }\href@noop {} {\bibfield  {journal}
  {\bibinfo  {journal} {Phys. Rev. B}\ }\textbf {\bibinfo {volume} {93}},\
  \bibinfo {pages} {035138} (\bibinfo {year} {2016})}\BibitemShut {NoStop}%
\bibitem [{\citenamefont {Skinner}\ \emph {et~al.}(2013)\citenamefont
  {Skinner}, \citenamefont {Chen},\ and\ \citenamefont
  {Shklovskii}}]{Skinner:TIsurface}%
  \BibitemOpen
  \bibfield  {author} {\bibinfo {author} {\bibfnamefont {Brian}\ \bibnamefont
  {Skinner}}, \bibinfo {author} {\bibfnamefont {Tianran}\ \bibnamefont {Chen}},
  \ and\ \bibinfo {author} {\bibfnamefont {B.~I.}\ \bibnamefont {Shklovskii}},\
  }\bibfield  {title} {\enquote {\bibinfo {title} {Effects of bulk charged
  impurities on the bulk and surface transport in three-dimensional topological
  insulators},}\ }\href@noop {} {\bibfield  {journal} {\bibinfo  {journal}
  {JETP}\ }\textbf {\bibinfo {volume} {117}},\ \bibinfo {pages} {579} (\bibinfo
  {year} {2013})}\BibitemShut {NoStop}%
\bibitem [{\citenamefont {Das~Sarma}\ \emph {et~al.}(2011)\citenamefont
  {Das~Sarma}, \citenamefont {Adam}, \citenamefont {Hwang},\ and\ \citenamefont
  {Rossi}}]{DasSarma:GrapheneReview}%
  \BibitemOpen
  \bibfield  {author} {\bibinfo {author} {\bibfnamefont {S.}~\bibnamefont
  {Das~Sarma}}, \bibinfo {author} {\bibfnamefont {Shaffique}\ \bibnamefont
  {Adam}}, \bibinfo {author} {\bibfnamefont {E.~H.}\ \bibnamefont {Hwang}}, \
  and\ \bibinfo {author} {\bibfnamefont {Enrico}\ \bibnamefont {Rossi}},\
  }\bibfield  {title} {\enquote {\bibinfo {title} {Electronic transport in
  two-dimensional graphene},}\ }\href@noop {} {\bibfield  {journal} {\bibinfo
  {journal} {Rev. Mod. Phys.}\ }\textbf {\bibinfo {volume} {83}},\ \bibinfo
  {pages} {407} (\bibinfo {year} {2011})}\BibitemShut {NoStop}%
\end{thebibliography}%

\newpage

\renewcommand{\theequation}{S\arabic{equation}}
\renewcommand{\thefigure}{S\arabic{figure}}
\renewcommand{\thetable}{S\arabic{table}}
\renewcommand{\thetable}{S\arabic{table}}
\renewcommand{\bibnumfmt}[1]{[S#1]}

\setcounter{equation}{0}
\setcounter{figure}{0}
\setcounter{enumiv}{0}

\onecolumngrid
\newpage
\begin{center}
\textbf{\large Supplemental Material for \\
``Electron transport in nodal-line semimetals''}
\end{center}
\vspace{2ex}
\twocolumngrid

\section*{Calculation of Transport and Elastic Scattering Times}
\label{sec:coefficient}

In what follows we present details of the calculation of the transport and elastic scattering times, $\tau_\text{tr}$ and 
$\tau_0$, respectively, in both regimes of low and high doping. The transport scattering time
for quasiparticles near a straight nodal line is given by
\begin{align}
\frac{1}{\tau_\text{tr}(\e)} & = &
\pi \Ni {\nu_{2D}}(\varepsilon) \int_{-\infty}^\infty
\frac{d k_z}{2\pi} \int_0^{2 \pi} \frac{d\theta}{2\pi}  (1 - \cos^2 \theta)
\nonumber\\
&  & \left| \phi\left[\left(k_z^2+\frac{4 \e^2}{v^2} \sin^2\frac{\theta}{2}\right)^\frac{1}{2}\right] \right|^2 ,
\label{eq:tautrSM}
\end{align}
which follows directly from Eq.~(\ref{TranspRateGen}). The quantity $\sqrt{k_z^2 + (4 \e^2/v^2) \sin^2 (\theta/2)}$ 
in Eq.~(\ref{eq:tautrSM})
is the change in quasiparticle momentum upon scattering by an angle $\theta$.  

The elastic scattering time is given by
\begin{align}
\frac{1}{\tau_{0}(\e)} & = &
\pi \Ni {\nu_{2D}}(kv) \int_{-\infty}^\infty
\frac{d k_z}{2\pi} \int_0^{2 \pi} \frac{d\theta}{2\pi}  (1 + \cos \theta)
\nonumber\\
&  & \left| \phi\left[\left(k_z^2+\frac{4 \e^2}{v^2}\sin^2\frac{\theta}{2}\right)^\frac{1}{2}\right] \right|^2 ,
\label{eq:tau0SM}
\end{align}
which differs from Eq.\ (\ref{eq:tautrSM}) only by a factor $(1 - \cos \theta)$ in the integrand.


\subsection{Low-doping regime}

The scattering rate of the quasiparticles at the Fermi energy and at low temperatures
depends on the details of the impurity potential at distances of order of the quasiparticle wavelength $v/|\mu|$
and, thus, on the polarisation operator at momenta of order of $k_F=|\mu|/v$.
Near each small straight segment of the nodal line the motion of electrons is effectively two-dimensional,
with the contribution to the polarisation operator given by the polarisation operator of $2D$ Dirac
electrons (graphene)\cite{Ando:polarisation,HwangSarma:GraphenePolarisation,DasSarma:GrapheneReview}
\be 
\Pi_\text{2D}(\bk) = \frac{g \kf}{2 \pi v} f\left( \frac{k}{2 \kf} \right),
\label{PiFunction}
\ee
multiplied by $d\kp/(2\pi)$, where $d\kp$ is the length of the segment;
$\bk$ is the two-dimensional momentum and $f(x)$ is a function that accounts for both interband and intraband electron polarisation, and is given by
\begin{align} 
f(x) = \left\{
	\begin{array}{cc}
		1, \quad & x < 1,
		\\
		1 + \frac{\pi x}{4} - \frac{1}{2} \sqrt{1 - x^{-2}} - \frac{x}{2} \arcsin(\frac{1}{x}), \quad & x > 1.
	\end{array}
	\right.
\end{align}

The polarisation function for the entire nodal line is obtained by adding up the contributions $\Pi_\text{2D}(\bk_\perp) \cdot d\kp/(2\pi)$ for each segment of length $d\kp$, where $\bk_\perp$ is the momentum perpendicular to the segment.  While the result of this addition depends in general on the shape of the nodal line, it will generally have the same functional form as $\Pi_\text{2D}(\bq)$ in terms of its dependence on the three-dimensional momentum $\bq$ for a given fixed direction of $\bq$.  In this paper we use the approximation of an isotropic polarisation function, so that
\be 
\Pi(\bq) = \text{const} \times \Pi_\text{2D}(q) \frac{\ko}{ 2\pi}.
\ee
In terms of the constant $C$ introduced in Eq.\ (\ref{PolarisationOperator}), the polarisation function can be written as
\be 
\Pi(\bq) = \frac{8 C g \kf \ko}{v} f \left( \frac{q}{2 \kf} \right).
\label{PiAlmostGraphene}
\ee

The screened potential is given, in terms of the polarisation operator, is given by
\be 
\phi(q) = \frac{\phi_0(q)}{1 - \Pi(q) \phi_0(q)}.
\ee
Using Eqs.~(\ref{eq:tautrSM}), (\ref{PiAlmostGraphene}) and (\ref{PiFunction}) and evaluating the integral numerically
yields Eq.\ (\ref{TransportTimeLow}) with $\gamma_\text{tr} \approx 175 C^2$.

The elastic scattering time $\tau_0$ can be evaluated similarly, leading to
 Eq.\ (\ref{TransportTimeLow}) with the constant $\gamma_\text{tr}$ replaced by the value $\gamma_\text{el} \approx 84.7 C^2$.

\subsection{High-doping regime}

In the high-doping regime, corresponding to strongly screened impurities,
the quasiparticle scattering is dominated by impurities far from the $x$-$y$ plane of the 
quasiparticle motion.  Since the potential of the impurities is screened exponentially, they contribute only small-momentum scattering, and therefore a typical scattering event rotates the quasiparticle momentum only by a small angle $\theta \ll 1$, as follows also directly from the integral (\ref{eq:tautrSM}).
Consequently, the transport scattering time $\tau_\text{tr}$ significantly exceeds the elastic scattering time $\tau_0$.  

One can evaluate $\tau_\text{tr}$ using Eq.\ (\ref{eq:tautrSM}) and the screened potential $\phi(q) = 4 \pi e^2/[\varkappa (q^2 + \ltf^{-2})]$, which is the Fourier transform of Eq.\ (\ref{Yukawa}).  This calculation gives the result announced in Eq.\ (\ref{TransportRateHigh}).  On the other hand, evaluating the elastic scattering rate using Eq.\ (\ref{eq:tau0SM}) gives
\be 
\tau_0 = \frac{1}{4 \pi \alpha^2 \Ni v \ltf^2},
\ee 
as announced in the main text.  This elastic scattering time is smaller than $\tau_\text{tr}$ by a factor $(\ln X) / X^2$, where $X = |\mu| \ltf/v \gg 1$ in the high-doping regime.

\end{document}